# Exploring the Physical Properties, Hydrogen Storage Capacity and Thermal Barrier Performance of $LaMg_2H_7$: A First-Principles Investigation

Tanvir Khan[1], Md Hasan Shahriar Rifat[2], M. Ibrahim[1], J.H. Abir[1], Saptak Roy[3], S.H. Naqib[1]

[1]Department of Physics, University of Rajshahi, Rajshahi 6205, Bangladesh
[2]Dept. of Materials Science and Engineering, University of Rajshahi, Rajshahi 6205, Bangladesh
[3]Department of Physics, Chittagong University of Engineering & Technology, Pahartali, Chittagong-4349, Bangladesh

## Abstract

$LaMg_2H_7$ is a ternary wide band gap semiconductor that is a member of the hydride family. The bulk physical properties of the $LaMg_2H_7$ compound, including its structural, electronic band structure, elastic, thermal, and optical properties, have been examined in this work using density functional theory (DFT). The elastic constants indicate that $LaMg_2H_7$ is mechanically stable, brittle in nature, and anisotropic. This compound possesses a moderate level of hardness. The band structure and density of states have been examined to have a better understanding of its electronic behavior. The intrinsic carrier concentrations and effective masses have been determined using the band structure. The gravimetric hydrogen storage capacity (Cwt%) has been calculated, indicating that this compound is suitable for hydrogen storage applications. This compound is dynamically stable, as confirmed by its phonon dispersion. Here, the reflectivity, absorption coefficient, refractive index, dielectric function, optical conductivity, and loss function of this wide-band-gap semiconductor are investigated. The substance is a moderate reflector of ultraviolet (UV) light. The absorption and conductivity support the gap in the band structure. The thermodynamic properties, such as bulk modulus, internal energy, specific heat capacity, entropy, thermal expansion coefficient, and Debye temperature, have been explored at varying temperatures and pressures. $LaMg_2H_7$ has a moderate level of melting temperature with higher lattice thermal conductivity. The value of the thermal expansion coefficient and minimum thermal conductivity is highly recommended for use as a thermal barrier coating (TBC).

# 1 Introduction

The need for energy is always rising as a result of the world's population growth. Much of this increasing energy consumption is still heavily reliant on fossil fuel reserves. This trend raises significant concerns about resource depletion and environmental sustainability [1]. The heavy dependence on fossil fuels to meet today's energy needs results in unchecked emissions of greenhouse gases. These emissions significantly contribute to air pollution and the acceleration of global warming. Identifying alternatives to fossil fuels is essential for a successful energy transition. Among the various options, hydrogen energy stands out as a promising solution. It is widely regarded as one of the most promising alternatives for meeting future energy demands sustainably. Hydrogen energy is gaining rapid attention as a promising solution to global energy challenges. It offers great promise for supporting the shift to a more sustainable future [2–4].

To identify a hydrogen storage system that satisfies the criteria of low weight for mobile applications and sufficient capacity, chemical hydrides based on lightweight metals have been extensively studied [5,6]. Magnesium (Mg) and its compounds have attracted significant attention as potential materials for hydrogen storage due to their favorable properties, including a high theoretical hydrogen storage capacity (7.6 wt%), lightweight nature, abundance in the Earth's crust, and low toxicity [7,8]. $MgH_2$, in particular, has been extensively studied owing to its ability to reversibly absorb and desorb hydrogen under moderate conditions. However, pure $MgH_2$ suffers from high thermodynamic stability and sluggish hydrogen sorption kinetics, which necessitate high temperatures (>300 °C) for desorption. To overcome these limitations, several strategies have been employed, including nanostructuring, alloying, and doping with transition metal (TM) catalysts to improve reaction kinetics and lower the activation energy [9,10]. These modifications enable Mg-based materials to become more competitive with other hydrogen storage media.

In comparison to Mg and its compounds, other lightweight metal clusters such as aluminum (Al) have also been explored for hydrogen storage. Al-based systems exhibit relatively good thermal stability and moderate hydrogen adsorption energies, though they generally offer lower hydrogen storage capacity than Mg-based counterparts [11]. Heavier noble metal clusters like copper ($Cu_n$), silver ($Ag_n$), and gold ($Au_n$) demonstrate stronger interactions with hydrogen molecules, often leading to higher adsorption energies but reduced reversibility and much lower gravimetric capacities due to their higher atomic weights [12,13]. Additionally, these noble metals are significantly more expensive and less abundant than Mg or Al. The introduction of TM dopants into these clusters has been shown to enhance hydrogen binding through d-orbital hybridization and electronic tuning, improving both adsorption energy and stability [8]. Nevertheless, the balance between adsorption strength and desorption capability remains a critical design factor, and Mg-based systems when modified appropriately continue to offer a compelling combination of high capacity, reversibility, and cost-effectiveness.

Carbon-based materials such as nanotubes, fullerenes, and other nanoporous structures have attracted significant attention as potential hydrogen storage media [14]. Interest in these materials surged after early work by Dillon et al., who reported hydrogen uptake levels as high as 5–10 wt% in single-walled carbon nanotubes under cryogenic conditions (133 K) and very low pressure (0.04 MPa) [15]. These findings, along with similarly impressive results for other carbon nanostructures [16], initially raised high expectations for their practical use in hydrogen storage. However, later studies reported much lower storage capacities around 1.55 wt% at 573 K and 0.01 MPa highlighting the limitations of these materials [17]. The reduced performance is largely due to the weak van der Waals forces governing the interaction between hydrogen molecules and carbon surfaces. Computational modeling has since become a valuable tool in exploring the hydrogen storage potential of two key carbon structures: planar graphite sheets with hexagonal patterns and single-walled carbon nanotubes, which are essentially rolled graphene sheets.

High-pressure synthesis has emerged as a powerful strategy for discovering new hydride compounds relevant to hydrogen storage. This technique is generally classified into two categories based on the pressure-generation mechanism: anvil-type equipment with solid media and autoclave-type apparatus with gas media. Notably, compounds such as $LaMg_2H_7$, $CeMg_2H_7$, and $CsMgH_3$ have been successfully synthesized under pressures of about 10 MPa using autoclave methods[18,19].

The $LaMg_2H_7$ compound is leaving a substantial gap in understanding its fundamental physical behavior. Crucially, no prior work has explored its mechanical robustness, anisotropy, thermophysical, thermodynamic, optical response, charge density or bonding nature all of which are vital for thermal barrier and optoelectronic applications. This study addresses these gaps by delivering the first systematic and multi-faceted investigation of $LaMg_2H_7$, correlating structure-property relationships with potential functional uses. The findings establish $LaMg_2H_7$ as a strong candidate for next-generation sustainable energy and device technologies.

.

# 2 Computational methodology

Here, density functional theory (DFT) was used to investigate the structural, electronic, elastic, thermophysical, and thermodynamic properties of the $LaMg_2H_7$ compound. Utilizing the full-potential linearized augmented plane wave plus local orbitals (FP-LAPW+lo) approach, the WIEN2k code was used to carry out the computations [20,21]. To accurately describe exchange and correlation effects, the GGA-PBEsol [22] method was employed here. For improved accuracy in estimating band gaps, especially important in semiconductors and perovskite materials, this study combines GGA-PBEsol with the modified Becke Johnson (mBJ) potential [23], known for producing results within approximately 2% of experimental values.

To reduce the number of plane waves and ensure the convergence of the calculation we have used 550 eV cutoff energy. For integration over the first Brillouin zone a 14×14×9 k-point mesh of Monkhorst-Pack scheme [24] was used. The elastic constants and phonons were calculated using CASTEP, but Thermodynamic properties were computed using Gibbs2 [25], which is already integrated into the WIEN2k package. Finally, VESTA was employed to visualize the crystal structures and analyze the polyhedral arrangements within the lattice, providing deeper insights into their geometric and bonding characteristics. Results and discussion

## 2.1 Crystal structure and hydrogen storage

A material's symmetry and crystal structure are important factors in influencing its physical characteristics, such as optical performance, electrical band structure, and elasticity. These properties are closely tied to the arrangement of atoms within the lattice, the distances between them, and the distribution of electronic states. In the case of $LaMg_2H_7$, the material crystallizes in a tetragonal structure with the space group $P41\overline{21}\overline{2}$ (No. 92). An illustration of this structure is shown in **Figure 1(a)**.

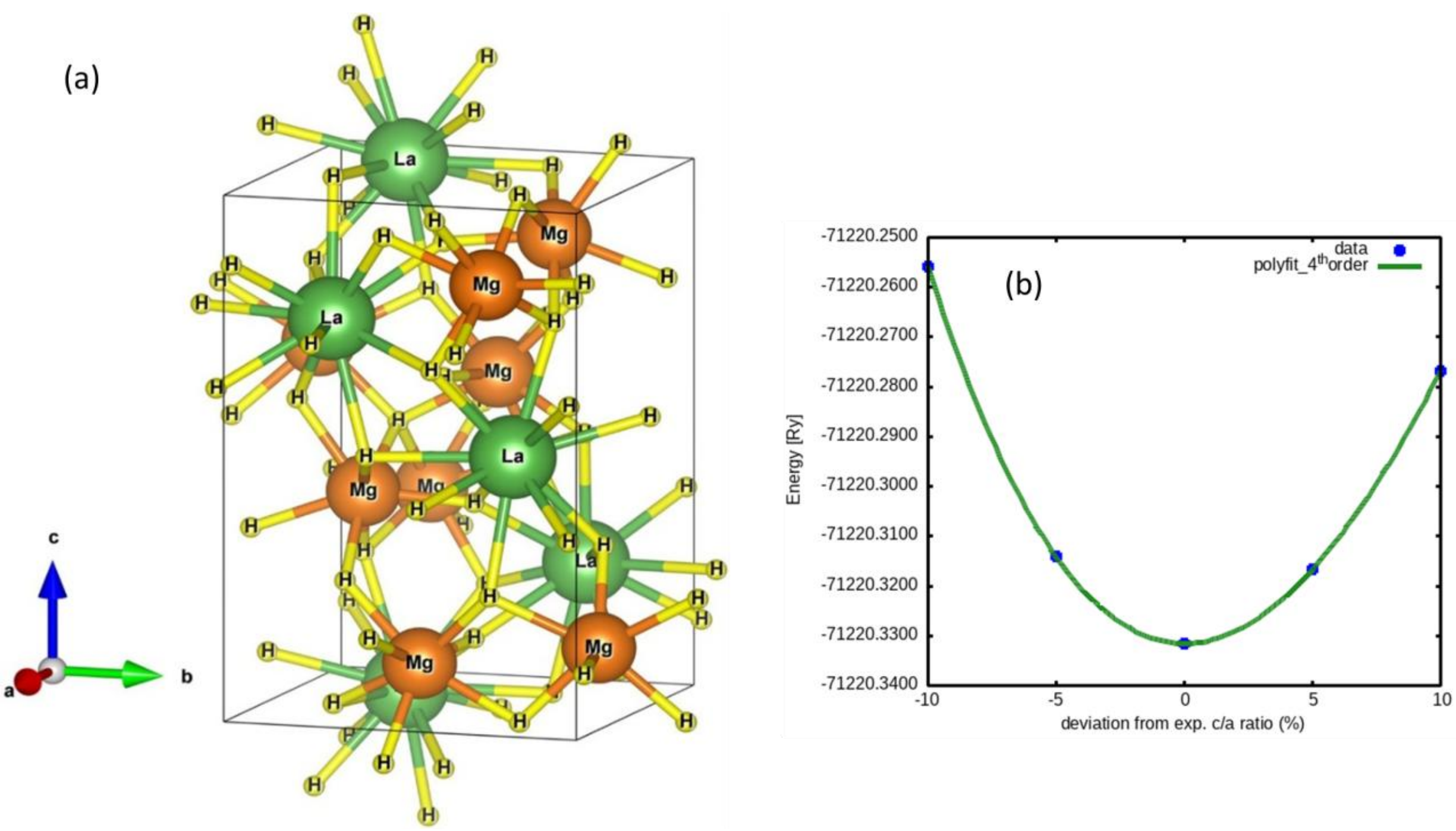

**Figure 1.** (a) crystal structure of $LaMg_2H_7$ and (b) Ground state energy of $LaMg_2H_7$ as a function of c/a ratio.

**Figure 1(b)** represents the variation of the ground styate energy of $LaMg_2H_7$ with the deviation of experimental c/a ratio. The minimum of the curve corresponds to the optimal (most stable) c/a ratio. Any deviation, either positive or negative, leads to an increase in energy, indicating reduced stability. The nearly symmetric nature of the curve around the minimum reflects a similar sensitivity of the system to both compression and expansion of the lattice ratio. The optimized lattice parameters and volume of $LaMg_2H_7$ are summarized in Table 1 along with the corresponding experimental values[13]. A comparison reveals that the calculated results are in reasonable agreement with the experimental data.The unit cell of $LaMg_2H_7$ contains four formula units. Each formula unit consists of one La atom, two Mg atoms, and seven H atoms. The details of the compound have been given to the ref. [26]. The potential of $LaMg_2H_7$ hydride perovskites for hydrogen storage applications has been evaluated by calculating their gravimetric hydrogen storage capacity. This was done using **Eq. 1**, where the amount of stored hydrogen is expressed in terms of the gravimetric weight percent ($C_{wt}$%). The algorithm used to determine the $C_{wt}$% follows a standard approach, as outlined in previous studies [27,28].

$$C_{wt} = \left( \frac{M_H \times \left(\frac{H}{M}\right)}{M_{Host} + \left(\frac{H}{M}\right) \times M_H} \right) \times 100\% \tag{1}$$

The weights of the hydrogen atoms ($M_H$), the material they are in ($M_{Host}$), and the ratio of hydrogen to material atoms (H/M) are all represented in **Eq. 1**.

**Table 1** illustrates the weight percentage of the gravimetric capacity of hydride $LaMg_2H_7$, which is 3.64%, indicating that this compound is quite suitable for hydrogen storage applications [29,30].

Formation energy ($\Delta E_f$) and cohesive energy ($\Delta E_c$), indicating the energy necessary to disrupt the bonds between various atoms in a crystal structure, was calculated for $LaMg_2H_7$ material using the following equations (**Eqs. 2 and 3)**

$$\boldsymbol{\Delta E_f}(\mathrm{LaMg_2H_7}) = \frac{\boldsymbol{E_{Total} - \left(4E_{gr}(La) + 8E_{gr}(Mg) + 28E_{gr}(H)\right)}}{\boldsymbol{40}} \tag{2}$$

$$\boldsymbol{\Delta E_c}(\mathrm{LaMg_2H_7}) = \frac{\boldsymbol{E_{Total} - (4E_{iso}(La) + 8E_{iso}(Mg) + 28E_{iso}(H))}}{\boldsymbol{40}} \tag{3}$$

Cohesive energy reflects the strength of atomic bonding within a solid, a more negative cohesive energy generally indicates greater structural stability. Formation enthalpy represents the energy change associated with the formation of a compound from its constituent elements. For hydride compounds, more negative formation enthalpy values suggest higher thermodynamic stability[45]. Our estimated vaues of $\Delta E_f$ and $\Delta E_c$ are shown in Table 1.**Table 1.** The optimized lattice parameters a=b (Å), c (Å), optimized cell volume V ( $Å^3$) formation energy **(eV/atom)** and cohesive energy **(eV/atom)** and the gravimetric weight percent $C_{wt}$ (%) of $LaMg_2H_7$ compound.

| Compound | a=b | c | V | Functionals | $\Delta E_f$ | $\Delta E_c$ | $C_{wt}$ | Ref. |
|---|---|---|---|---|---|---|---|---|
| $LaMg_2H_7$ | 6.4054 | 9.5994 | 383.040 | - | - | - | - | [26] |
| | 6.3370 | 9.5360 | 383.041 | GGA-PBEsol | -0.847 | -3.73 | 3.64 | This work |

## 2.2 Elastic properties and anisotropy

To understand the mechanical behavior of solid materials, it is essential to determine their independent elastic constants. Elastic constants provide a comprehensive description of how solids respond to external stresses and strains, as well as insights into the nature of atomic bonding and cohesive energy. They are directly linked to fundamental mechanical properties such as ductility, stability, brittleness, stiffness, and elastic anisotropy[31–34]. [31–34]. **Table 2** lists the elastic constants of $LaMg_2H_7$. $LaMg_2H_7$ has six elastic constants ($C_{11}$, $C_{12}$, $C_{13}$, $C_{33}$, $C_{44}$, and $C_{66}$) because of their tetragonal structure. Mechanical stability must be veryfied using the Born criteria, [35] which for tetragonal systems are given by the conditions in **Eq. 4.**

$$C_{11} > 0, C_{33} > 0, C_{44} > 0, C_{66} > 0, C_{11} + C_{33} - 2C_{13} > 0,$$
$$C_{11} - C_{12} > 0, 2(C_{11} + C_{12}) + C_{33} + C_{44} > 0. \quad (4)$$

Here, the calculated values satisfy all the criteria, indicating that the structure is mechanically stable. The elastic constants $C_{11}$ and $C_{33}$ represent the stiffness against uniaxial deformation along [100] and [001] crystallographic directions. According to Table 2, $C_{11}$ is greater than $C_{33}$, which means that the material is stiffer along the [100] direction.

**Table 2**. Calculated elastic constants $C_{ij}$ (in GPa) of $LaMg_2H_7$.

| **Compound** | **$C_{11}$** | **$C_{12}$** | **$C_{13}$** | **$C_{33}$** | **$C_{44}$** | **$C_{66}$** | **Ref.** |
|---|---|---|---|---|---|---|---|
| $LaMg_2H_7$ | 131.88 | 43.39 | 34.14 | 129.59 | 46.04 | 62.86 | This work |
| $LaH_2$ | 90.3 | 49.4 | --- | --- | 47.5 | --- | [36] |
| **$K_2MgH_4$** | 47.05 | 12.84 | 11.39 | 38.27 | 15.38 | 26.68 | [37] |

Consequently, $LaMg_2H_7$ is more easily deformed and mechanically softer along the [001] direction Similar behavior is shown in $K_2MgH_4$. To determine the value of bulk modulus (B), shear modulus (G), Young's modulus (Y), and Poisson's ratio ($v$), the following equations (**Eqs. 5-12**) have been used, which use the Voigt–Reuss [38,39] approximation**.**

$$B_V = \frac{2}{9}\left(C_{11} + C_{12} + 2C_{13} + \frac{C_{33}}{2}\right) \quad (5)$$

$$B_R = C^2/M; \text{ where } C^2 = (C_{11} + C_{12})C_{33} - 2C_{13}^2 \text{ and } M = C_{11} + C_{12} + 2C_{33} - 4C_{13} \quad (6)$$

$$G_V = \frac{M + 3C_{11} - 3C_{12} + 12C_{44} + 6C_{66}}{30} \quad (7)$$

$$G_R = \frac{15}{\frac{18B_V}{C^2} + \frac{6}{(C_{11} - C_{12})} + \frac{6}{C_{44}} + \frac{6}{C_{66}}} \quad (8)$$

$$B = \frac{B_V + B_R}{2} \quad (9)$$

$$G = \frac{G_V + G_R}{2} \quad (10)$$

$$E = \frac{9BG}{(3B + G)} \quad (11)$$

$$v = \frac{3B - E}{6B} \quad (12)$$

The calculated bulk modulus (B) of the studied hydride is 68.5 GPa, which is higher than those of other reported hydrides such as $LaH_2$ (63 GPa) and $K_2MgH_4$ (22.5 GPa). This indicates that the present hydride is relatively stiffer compared to these compounds. The relative magnitudes of the elastic moduli indicate that the material is more rigid under uniaxial tension, as reflected by Young's modulus (Y) being greater than B, and B in turn being greater than the shear modulus (G). This implies that the material is less rigid under uniform volume changes (B), while it is softest and deforms most readily under shear stress (G).The ratio of B and G is known as Pugh's ratio; this ratio provides two solid characteristics, ductile and brittle [40]. A material is called ductile if B/G > 1.75; otherwise, it is brittle in nature. From **Table 3,** the value of B/G is 1.39, so according to the above discussion this material is brittle, which is consistent with the hydride $K_2MgH_4$.

**Table 3**. Calculated bulk modulus *B* (GPa), shear modulus *G* (GPa), Young modulus *Y* (GPa), Poisson's ratio *v*, Pugh's ratio B/G, elastic anisotropic factor *A,* and universal elastic constant $A^U$ of $LaMg_2H_7$.

| Compound | *B* | *G* | *Y* | *v* | *B/G* | *A* | $A^U$ | Ref. |
|---|---|---|---|---|---|---|---|---|
| $LaMg_2H_7$ | 68.5 | 49.4 | 119.4 | 0.209 | 1.39 | 0.95 | 0.008 | This work |
| $LaH_2$ | 63.0 | --- | --- | --- | --- | --- | --- | [36] |
| $K_2MgH_4$ | 22.5 | 17.5 | 41.7 | 0.19 | 1.28 | --- | --- | [37] |

Poisson's ratio, *v,* can be used to determine the covalent or ionic bonds of a material. The value of *v* for covalent material is usually 0.10, and for ionic material it is 0.25 [41]. For $LaMg_2H_7$, the value of *v* is 0.209, which suggests that ~~is~~ ionic contribution dominates its bonding nature. A similar bonding character is also observed in $K_2MgH_4$.Elastic anisotropy has a profound influence on various physical processes, including dislocation dynamics, plastic deformation in solids, internal abrasion, and the appearance and propagation of fractures [37]. The elastic anisotropy factor for a tetragonal structure along the [100] direction is determined by **Eq. 13.**

$$A = \frac{4C_{44}}{C_{11} + C_{33} - 2C_{13}} \tag{13}$$

For a tetragonal system, the elastic anisotropy factor equals 1 for isotropic behavior; values less than or greater than 1 indicate anisotropy. The calculated value for $LaMg_2H_7$ is 0.95, so this material is anisotropic. The universal anisotropic constant can be determined by **Eq. 14** [42].

$$A^U = \frac{5G_V}{G_R} + \frac{B_V}{B_R} - 6 \tag{14}$$

$A^U$ indicates whether the material is isotropic or anisotropic. . If the value of $A^U$ is exactly zero, the material is considered to be isotropic; otherwise, it is anisotropic.

Hardness refers to a material's ability to resist permanent deformation, which is one of the useful mechanical properties of materials that excel in industrial applications. Anti-friction coatings, heavy-duty components, low-emissivity glass coatings, optoelectronic devices, and microelectronics are all included in these application areas [43,44]. There is a relation between the material's strength and hardness [45]. Different frameworks have been proposed in the literature [60-64]. We have calculated the Vickers hardness $(H_V)_{micro}$, $(H_V)_{macro}$, $(H_V)_{Tian}$, $(H_V)_{Tetra}$, and $(H_V)_{Mazhnik}$ of $LaMg_2H_7$ by using **Eqs. 15-19**. In **Eq. 19**, $\chi(\nu)$ is a function of Poisson's ratio where $\gamma_0$ is a constant with the value of 0.096 which is dimensionless and $\chi(\nu)$ is represented by **Eq. 20**.

$$(H_V)_{micro} = \frac{(1-2\nu)Y}{6(1+\nu)} \tag{15}$$

$$H_V = 2\left[\left(\frac{G}{B}\right)^2 G\right]^{0.585} - 3 \tag{16}$$

$$(H_V)_{Tian} = 0.92\left(\frac{G}{B}\right)^{1.137} G^{0.708} \tag{17}$$

$$(H_V)_{Teter} = 0.151G \tag{18}$$

$$(H_V)_{Mazhnik} = \gamma_0 \chi(\nu) Y \tag{19}$$

$$\chi(\nu) = \frac{1 - 8.5\nu + 19.5\nu^2}{1 - 7.5\nu + 12.2\nu^2 + 19.6\nu^3} \tag{20}$$

The value of hardness is given in **Table 4,** from which it is clear that $LaMg_2H_7$ material appears to be reasonably hard. Hence, the variation in hardness values obtained from different methods seems justifiable. Notably, the calculated $(H_V)_{micro}$ and $(H_V)_{macro}$ values are 9.09 GPa and 10.35 GPa, respectively, indicating that the studied compound exhibits a moderate level of hardness.

**Table 4.** Calculated Vickers hardness (GPa) of $LaMg_2H_7$ using elastic moduli and Poisson's ratio.

| **Compound** | ***(HV)micro*** | | | | | **Ref.** |
|---|---|---|---|---|---|---|

| **Compound** | $(H_V)_{micro}$ | $(H_V)_{macro}$ | $(H_V)_{Tian}$ | $(H_V)_{Tetra}$ | $(H_V)_{Mazhnik}$ | **Ref.** |
|---|---|---|---|---|---|---|
| $LaMg_2H_7$ | 9.09 | 10.35 | 10.02 | 7.45 | 5.98 | This work |

The fluctuation of Young's modulus (Y), shear modulus (G), Poisson's ratio (σ), and compressibility was investigated using the ELATE [46] code. Here, the anisotropic nature of the material is observed. For the isotropic crystal, the 3D plots are spherical in shape. The deviation from perfect sphericity indicates anisotropy. For all properties, 2D and 3D plots are displayed in **Figure 2(a-d)**. Among the properties analyzed, only compressibility exhibits a fully spherical shape, while the others show deviations from their respective planes.

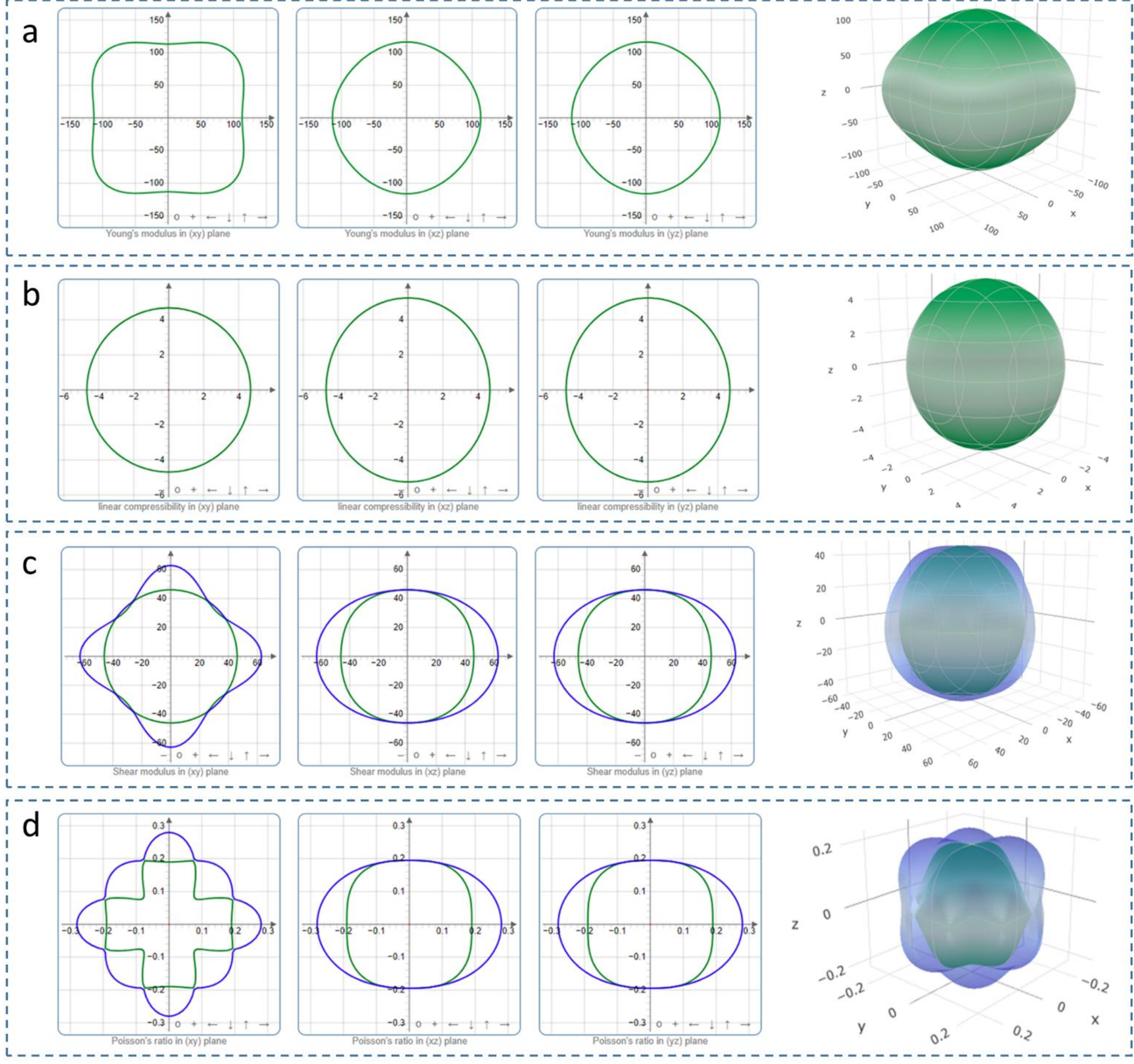

**Figure 2**. 2D and 3D plots of (a) Young's modulus, (b) linear compressibility, (c) shear modulus, and (d) Poisson's ratio.

## 2.3 Electronic band structure and density of states

Analyzing a material's electronic band structure helps us to understand a number of important phenomena, such as chemical bonding, electronic transport, superconductivity, optical characteristics, and magnetism. In particular, the bands at the Fermi level play a dominant role in shaping its electronic behavior. Accurately modeling devices and nanostructures also requires knowledge of effective charge carrier masses and well-defined band shapes. This understanding is essential across many areas of materials research, including catalyst design [47], rapid discovery of new battery materials [48], magnetism, and the development of superconductors [49].

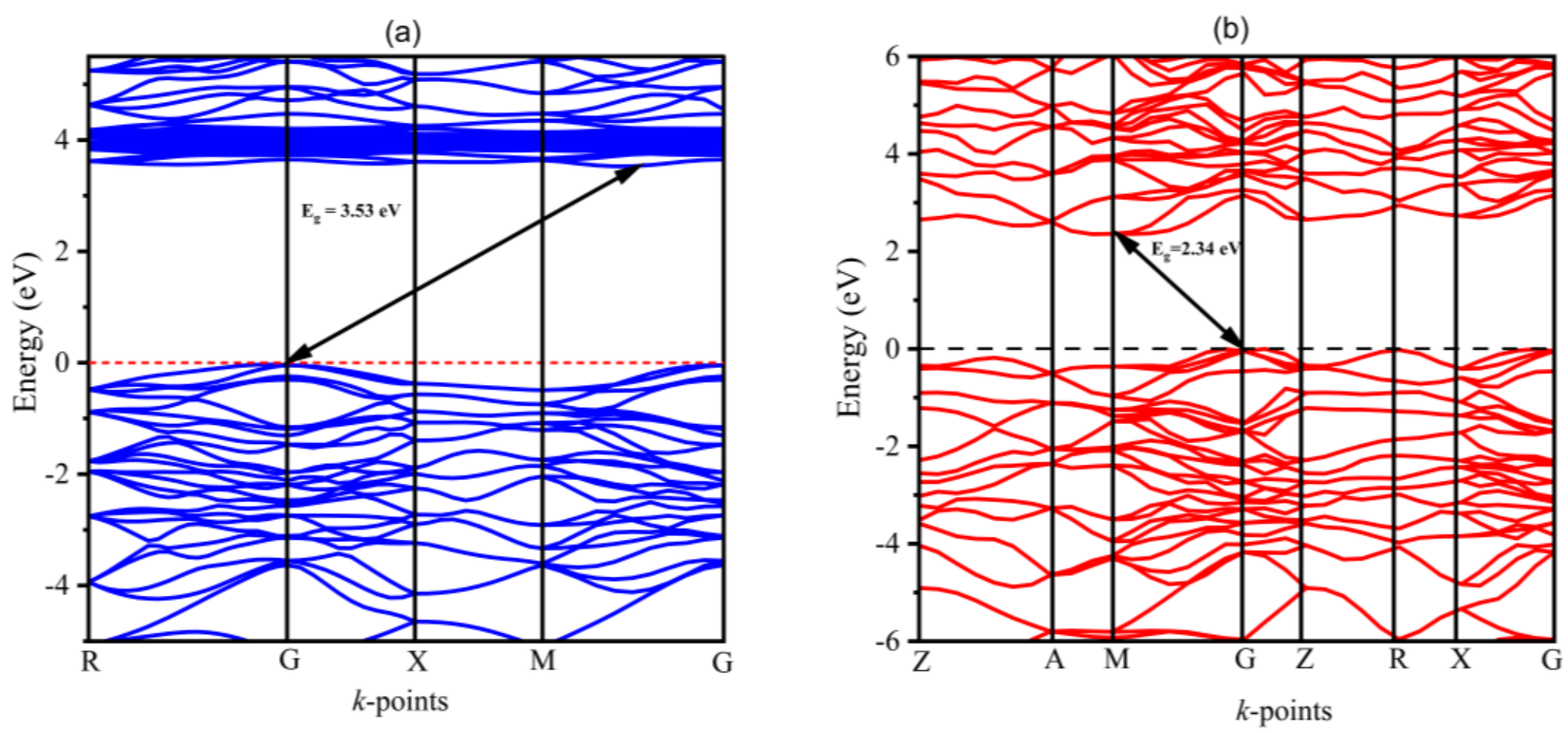

**Figure 3.** (a) band structure of $LaMg_2H_7$ compound (Using Wien2K- mBJ) (b) band structure of $LaMg_2H_7$ compound (using CASTEP-GGA).

The electronic band structures of LaMg2H7 compound were calculated along high-symmetry directions of the first Brillouin zone: Z–A–M–Γ–Z–R–X–Γ using CASTEP (GGA-PBEsol) and R–Γ–X–M–Γ using Wien2k (mBJ), over an energy range of –6 to +6 eV, as shown in Figures 3(a) and 3(b), respectively. As the minimum of the conduction band and the maximum of the valence band lies in different k-vectors, $LaMg_2H_7$ exhibits indirect band gaps in both case The values of the band gaps are 2.34 eV using CASTEP (GGA-PBEsol) and 3.53 eV using Wien2k. This indicates that the GGA functional underestimates the band gap. To improve accuracy, the modified Becke-Johnson (mBJ) method was employed in Wien2k. The electronic patterns of band structure of LaMg2H7 demonstrate that the energy bands show the varying degree of dispersion.

The effective masses of electrons and holes, as well as the carrier concentrations in the conduction and valence bands, were calculated. Additionally, the intrinsic carrier concentration was evaluated, all derived using standard equations. For improved accuracy, the modified Beck-Johnson approach (mBJ) [23] method was employed to calculate the band gap in Wien2k. The effective masses for electrons and holes have been computed, as has the carrier concentration of both electrons and holes in the conduction and valence bands. The intrinsic carrier concentration has also been explored. These were derived using the standard equations **(Eqs. 21-24)** [50].

$$\frac{1}{m*} = \frac{1}{\hbar}\frac{\partial^2 E(k)}{\partial K^2} \tag{21}$$

$$N_c = 2\left(\frac{2\pi kTm_e^*}{h^2}\right)^{\frac{3}{2}} \tag{22}$$

$$N_v = 2\left(\frac{2\pi kTm_h^*}{h^2}\right)^{\frac{3}{2}} \tag{23}$$

$$n_i = \sqrt{N_c N_v}\, exp\left(-\frac{E_g}{2kT}\right) \tag{24}$$

where E(k) is the energy as a function of the wave vector k, and ħ is the reduced Planck constant. This parabolic approximation around the band extrema enables the estimation of effective masses [51]. $N_c$ and $N_v$ are the carrier concentrations of electrons and holes. The $n_i$ indicates intrinsic carrier concentration, and $E_g$ is the band gap. Here, k is the Boltzmann constant and T is the room temperature.

**Table 5** lists the resulting effective masses (in units of the free electron mass $m_0$), Carrier concentration for both electrons and holes, and the intrinsic carrier concentration of $LaMg_2H_7$ compound. As shown in **Figure 3**, flatter bands correspond to heavier carriers with higher localization, whereas more dispersive bands give rise to lighter, more mobile charge carriers.

**Table 5.** Calculation of the effective mass of electrons ($m_e^*$), holes ($m_h^*$), electron carrier concentration $N_c$ ($cm^{-3}$), hole carrier concentrations $N_v$ ($cm^{-3}$) and intrinsic carrier concentrations $n_i$ ($cm^{-3}$).

| $m_e^*$ | $m_h^*$ | $N_c \times 10^{18}$ | $N_v \times 10^{18}$ | $n_i \times 10^{-11}$ |
|---|---|---|---|---|
| $0.26858m_0$ | - $0.36287m_0$ | 3.48 | 5.46 | 2.00 |

The electronic band structure governs key transport properties, such as carrier effective masses, which directly influence the performance of photovoltaic materials. A low carrier effective mass indicates high charge carrier mobility and efficient hole transport, both of which are essential for solar cell applications[52]. Moreover, the dielectric constant of a material is closely related to its electronic structure; a higher dielectric constant suppresses charge carrier recombination, thereby enhancing device efficiency[53]. Importantly, the optical properties of solids, including fundamental absorption processes, are also determined by the band structure. When photons interact with the material, optical absorption arises either from excitonic effects or from direct band-to-band electronic transitions, making the band gap a critical parameter that links electronic and optical behavior. [54]. At this absorption edge, two transition types are possible: direct and indirect. In both, a photon excites an electron from the valence band into the conduction band across the fundamental bandgap [55,56]. Near the band edge, this typically means an electron at the top of the valence band is promoted to the bottom of the conduction band [57]. However, if the electrons encounter structural or compositional disorder during this process, the density of states ρ($h\nu$) (where $h\nu$ is the photon energy) develops a "tail" that extends into the bandgap. This feature, known as the Urbach tail, causes the absorption coefficient $\alpha(h\nu)$ to decrease exponentially with photon energy. The characteristic energy associated with this tail, called the Urbach energy, can be evaluated from **Eq. 25**

$$\alpha(h\nu) = \alpha_0 \exp\left(\frac{h\nu}{E_u}\right) \tag{25}$$

Where $\alpha_o$ is a constant, '$h\nu$ ' is the photon energy and $E_u$ is the Urbach energy [58,58].

**Table 5** indicates that the intrinsic carrier concentration in $LaMg_2H_7$ is extremely low due to its wide bandgap. The density of states (DOS) provides information on the number of available electronic states at each energy level within the material, expressed per unit energy and volume. A higher density of states at a given energy level implies that more electronic states can be occupied. To better understand the electronic properties of $LaMg_2H_7$, both the total density of states (TDOS) and the partial densities of states (PDOS) calculations were performed, as illustrated in **Figure 4**. The analysis reveals crucial insights into the electronic behavior of the compound. The Fermi level is indicated by a vertical dashed line at zero energy, serving as a reference point for the energy levels of the electronic states.

The absence of any finite value of DOS at the Fermi level confirms the nonmetallic character of the materials. This observation aligns well with the band structure, further supporting the semiconducting nature of the compounds. The band gap observed in the TDOS closely matches that in the band structure, reinforcing the classification of the material as a wide-gap semiconductor. The partial density of states (PDOS) helps to understand which atomic orbitals contribute to the valence and conduction bands. For the compound, the top of the valence band is primarily hybridized of Hydrogen s-orbital and Lanthanum d-orbital, suggesting the presence of covalent bonding within the crystal structure. Meanwhile, the conduction band minimum is also largely influenced by La-d orbitals, with a minor contribution from Mg-p orbitals due to partial hybridization. The partial density of states (PDOS) of $LaMg_2H_7$ also reveals that the valence band is predominantly composed of hydrogen 1s states, strongly hybridized with magnesium 3s and a smaller contribution from Mg 3p orbitals, indicating an sp-like mixing on Mg that forms σ-type Mg–H bonds. This strong H–Mg hybridization accounts for the covalent character within the [$MgH_7$] units. In contrast, lanthanum contributes very little to the valence band, while the conduction band minimum is mainly derived from La 5d states, with minor contributions from La 6s/6p and narrow La 4f states appearing at higher energies. The clear separation between the filled H–Mg bonding states and the empty La-centered conduction states produces the wide band gap, consistent with the compound's semiconducting natuture.

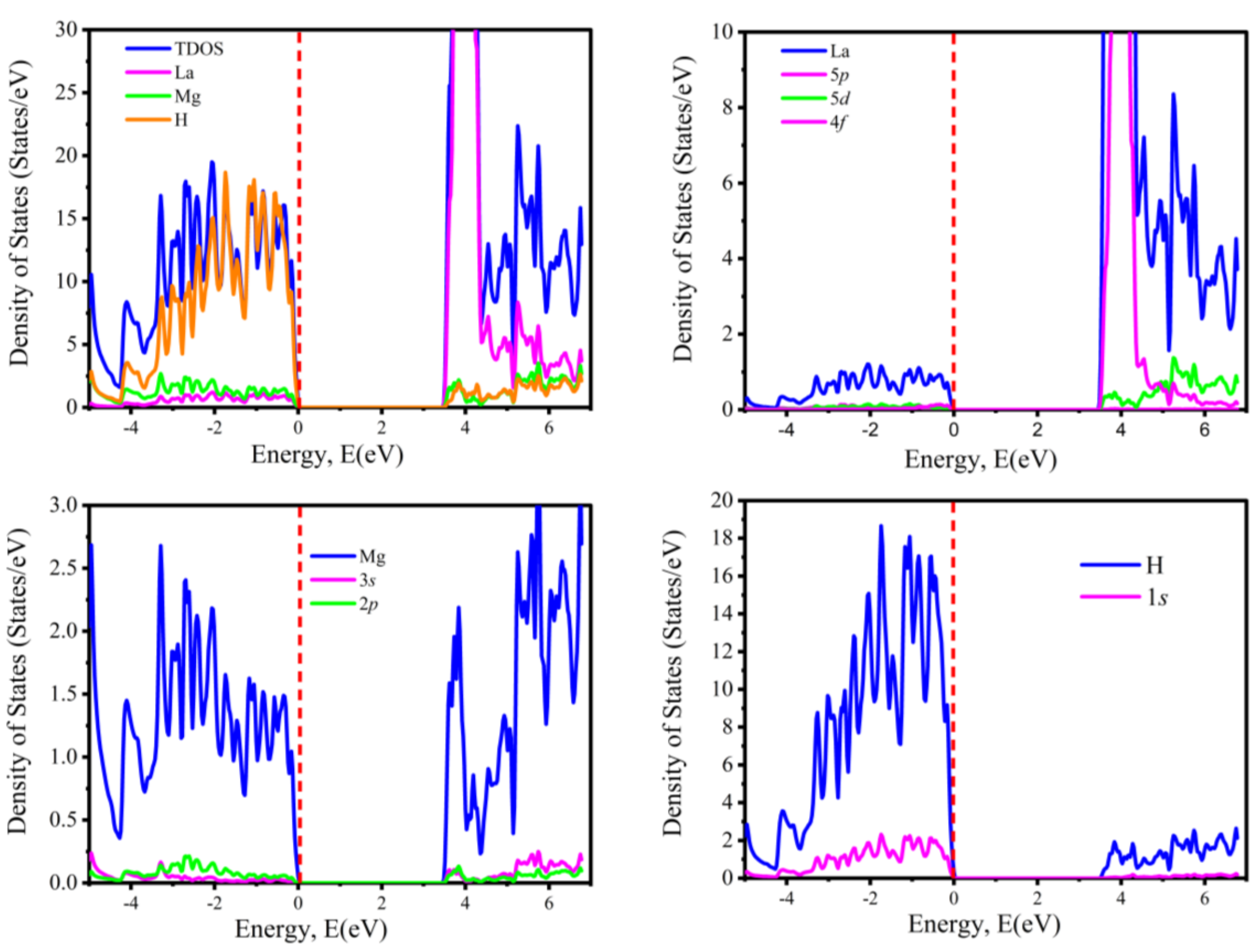

**Figure 4.** Total density of states (TDOS) and partial density of states (PDOS) of $LaMg_2H_7$ is plotted using the Wien2k package.

### *2.3.1 Mulliken atomic and bond overlap population*

Mulliken population analysis is used to investigate the properties of chemical bonds and effective valence charge (EVC) that provide multiple functional explanations for the distribution of electrons among the different bond components. In order to analyze the population in the CASTEP code, Sanchez-Portal established a method that uses projection of the plane wave function that assigns charges in linear combination of atomic orbital (LCAO) basis sets [59].

The Mulliken charge assigned to a certain atomic species a can be evaluated **Eqs. 26-27**.

$$Q(\alpha) = \sum_{k} w_k \sum_{\mu}^{on\ \alpha} \sum_{\nu} P_{\mu\upsilon}(k)\, S_{\mu\upsilon}(K) \tag{26}$$

$$P(\alpha\beta) = \sum_{k} w_k \sum_{\mu}^{on\ \alpha} \sum_{\nu}^{on\ \beta} 2P_{\mu\upsilon}(k)\, S_{\mu\upsilon}(K) \tag{27}$$

where $P_{\mu\nu}$ denotes the density matrix elements and $S_{\mu\nu}$ refers the overlap matrix. The results for three investigated structures based on electronic chages are shown in **Table 6.**

The bonding nature in $LaMg_2H_7$ can be interpreted through the Mulliken population analysis and effective valence charge distribution. The hydrogen atoms exhibit negative Mulliken charges in the range of –0.38 to –0.50, which indicates their electron-accepting character. The corresponding effective valence charges of 0.50–0.62 further confirm that H behaves as an anionic species. Magnesium atoms, on the other hand, carry small positive Mulliken charges of about +0.63 e with effective valence charges of 1.37, suggesting a partial donation of electrons. This reflects a mixed ionic–covalent interaction between Mg and H, as magnesium contributes electron density while still maintaining significant covalent overlap with hydrogen.

Lanthanum atoms show a much larger positive Mulliken charge (+1.60 e) and effective valence charge (1.40), highlighting their strong tendency to lose electrons and reinforcing the ionic character of La–H interactions. The presence of La in the compound therefore enhances the ionic contribution to the overall bonding. The variation in Mulliken charges among the hydrogen sites (–0.38 to –0.50) also indicates non-equivalent hydrogen environments, where some H atoms participate in stronger covalent interactions with Mg while others are more ionically bonded with La.

**Table 6:** Charge spilling parameter (%), orbital charges (electron), atomic Mulliken charges (electron), and effective valence charge (electron) of $LaMg_2H_7$.

| Compound | Charge spilling | Species | Mulliken atomic population | | | | Mulliken charge | Formal ionic charge | Effective Valence charge |
|---|---|---|---|---|---|---|---|---|---|
| | | | s | p | d | Total | | | |
| $LaMg_2H_7$ | 0.13 | | 1.38 | 0 | 0 | 1.38 | -0.38 | -1 | 0.62 |
| | | H | 1.41 | 0 | 0 | 1.41 | -0.41 | -1 | 0.59 |
| | | H | 1.38 | 0 | 0 | 1.38 | -0.38 | -1 | 0.62 |
| | | H | 1.38 | 0 | 0 | 1.38 | -0.38 | -1 | 0.62 |
| | | H | 1.41 | 0 | 0 | 1.41 | -0.41 | -1 | 0.59 |
| | | H | 1.38 | 0 | 0 | 1.38 | -0.38 | -1 | 0.62 |
| | | H | 1.38 | 0 | 0 | 1.38 | -0.38 | -1 | 0.62 |
| | | H | 1.41 | 0 | 0 | 1.41 | -0.41 | -1 | 0.59 |
| | | H | 1.38 | 0 | 0 | 1.38 | -0.38 | -1 | 0.62 |
| | | H | 1.38 | 0 | 0 | 1.38 | -0.38 | -1 | 0.62 |
| | | H | 1.41 | 0 | 0 | 1.41 | -0.41 | -1 | 0.59 |
| | | H | 1.38 | 0 | 0 | 1.38 | -0.38 | -1 | 0.62 |
| | | H | 1.38 | 0 | 0 | 1.38 | -0.38 | -1 | 0.62 |
| | | H | 1.41 | 0 | 0 | 1.41 | -0.41 | -1 | 0.59 |
| | | H | 1.38 | 0 | 0 | 1.38 | -0.38 | -1 | 0.62 |
| | | H | 1.38 | 0 | 0 | 1.38 | -0.38 | -1 | 0.62 |
| | | H | 1.41 | 0 | 0 | 1.41 | -0.41 | -1 | 0.59 |
| | | H | 1.38 | 0 | 0 | 1.38 | -0.38 | -1 | 0.62 |
| | | H | 1.38 | 0 | 0 | 1.38 | -0.38 | -1 | 0.62 |
| | | H | 1.41 | 0 | 0 | 1.41 | -0.41 | -1 | 0.59 |
| | | H | 1.38 | 0 | 0 | 1.38 | -0.38 | -1 | 0.62 |
| | | H | 1.38 | 0 | 0 | 1.38 | -0.38 | -1 | 0.62 |
| | | H | 1.41 | 0 | 0 | 1.41 | -0.41 | -1 | 0.59 |
| | | H | 1.38 | 0 | 0 | 1.38 | -0.38 | -1 | 0.62 |
| | | H | 1.50 | 0 | 0 | 1.50 | -0.50 | -1 | 0.50 |
| | | H | 1.50 | 0 | 0 | 1.50 | -0.50 | -1 | 0.50 |
| | | H | 1.50 | 0 | 0 | 1.50 | -0.50 | -1 | 0.50 |
| | | H | 1.50 | 0 | 0 | 1.50 | -0.50 | -1 | 0.50 |
| | | Mg | 2.56 | 6.28 | 0 | 9.37 | 0.63 | 2 | 1.37 |
| | | Mg | 2.56 | 6.28 | 0 | 9.37 | 0.63 | 2 | 1.37 |
| | | Mg | 2.56 | 6.28 | 0 | 9.37 | 0.63 | 2 | 1.37 |
| | | Mg | 2.56 | 6.28 | 0 | 9.37 | 0.63 | 2 | 1.37 |
| | | Mg | 2.56 | 6.28 | 0 | 9.37 | 0.63 | 2 | 1.37 |
| | | Mg | 2.56 | 6.28 | 0 | 9.37 | 0.63 | 2 | 1.37 |
| | | Mg | 2.56 | 6.28 | 0 | 9.37 | 0.63 | 2 | 1.37 |
| | | Mg | 2.56 | 6.28 | 0 | 9.37 | 0.63 | 2 | 1.37 |
| | | La | 1.76 | 5.85 | 1.80 | 9.40 | 1.60 | 3 | 1.40 |
| | | La | 1.76 | 5.85 | 1.80 | 9.40 | 1.60 | 3 | 1.40 |
| | | La | 1.76 | 5.85 | 1.80 | 9.40 | 1.60 | 3 | 1.40 |
| | | La | 1.76 | 5.85 | 1.80 | 9.40 | 1.60 | 3 | 1.40 |

### *2.3.2 Charge density*

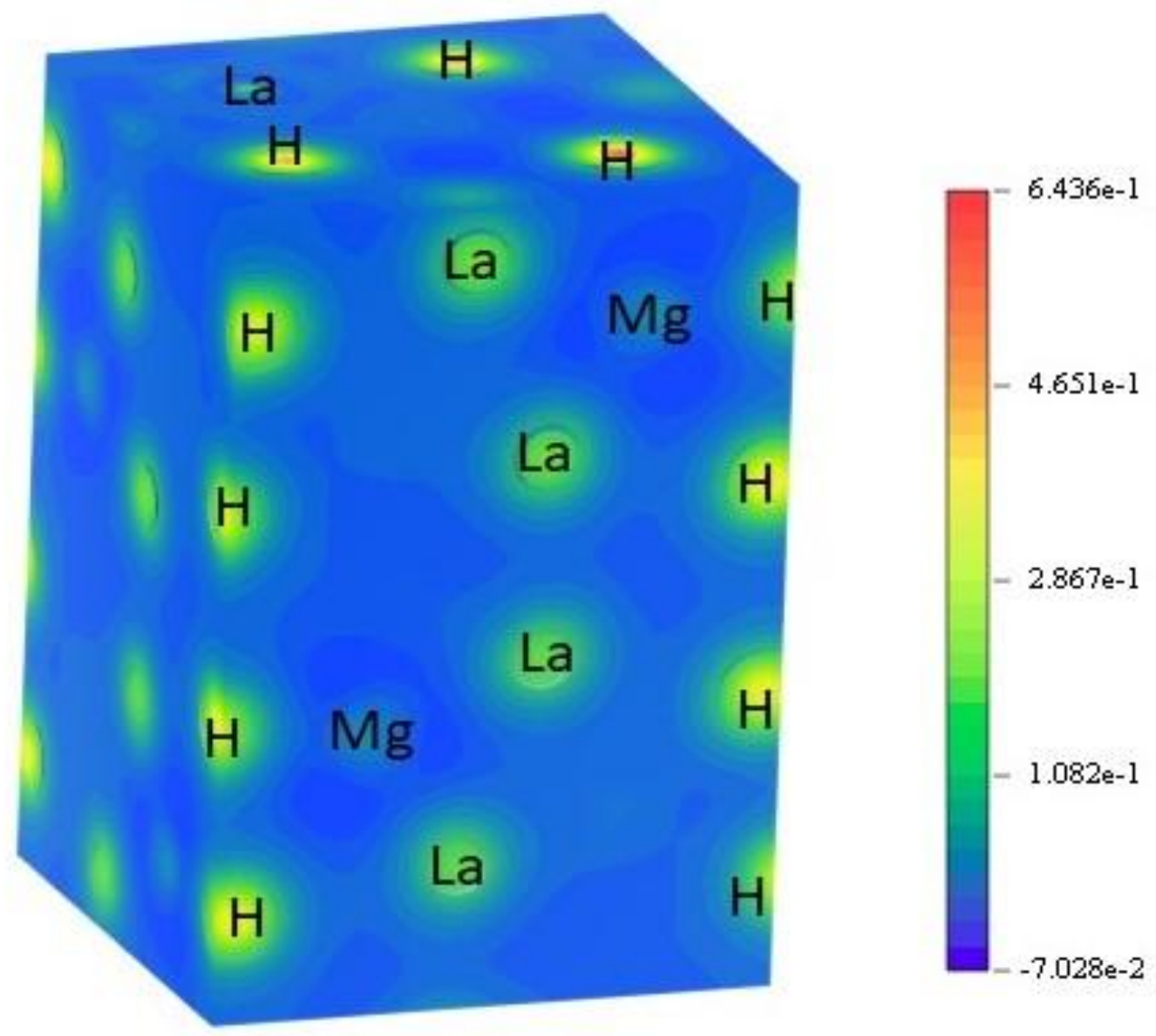

Figure 5. charge density of $LaMg_2H_7$ compound

The charge density contours of $LaMg_2H_7$ show significant electron accumulation around H atoms, while La and Mg sites possess relatively lower electron density. This indicates charge transfer from electropositive La and Mg atoms toward hydrogen, confirming the predominantly ionic nature of bonding. A slight overlap of charge contours between Mg and H suggests weak covalent hybridization, leading to mixed ionic-covalent bonding characteristics.

Mulliken analysis shows that H atoms gain electrons (−0.38 to −0.50 e), while Mg (+0.63 e) and La (+1.60 e) donate charge, confirming ionic bonding. Deviations from formal ionic charges and orbital hybridization indicate a small covalent contribution, consistent with the charge density results.

## 2.4 Optical properties

The optical properties of a material are determined how its charge carriers interact with incident photons or electromagnetic waves at the material's surface. In optoelectronic devices, particular attention is given to how materials respond to visible light. The study of solid materials using optical techniques has sparked widespread interest in modern science and technology. Optical materials play a vital role in a range of applications, including reconfigurable photonics, solar cells, lasers, photodetectors, sensors, and display technologies [60]. A material's optical behavior determines how it interacts with electromagnetic radiation. In most cases, the focus is on how optoelectronic devices respond to visible light. These investigations help identify materials that are well-suited for specific optoelectronic applications. These studies help predict which materials are best suited for different optoelectronic applications. Key optical properties that are typically analyzed as functions of photon energy include the dielectric function ε(ω), refractive index n(ω), optical conductivity σ(ω), reflectivity R(ω), absorption coefficient α(ω), and energy loss function L(ω), where ω (omega) is the angular frequency, defined as 2πf. In this section, we explore how these parameters respond to incident photon energy. The complex dielectric function is expressed as follows in (**Eqs. 28-34**).

$$\varepsilon(\omega) = \epsilon_1(\omega) + i\epsilon_2(\omega) \tag{28}$$

Kramers-Kronig relationships connect the imaginary and real parts. All additional optical constants of importance are given by the following relationships [61].

$$n(\omega) = \sqrt{\frac{|\varepsilon(\omega)| + \epsilon_1(\omega)}{2}} \tag{29}$$

$$k(\omega) = \sqrt{\frac{|\varepsilon(\omega)| - \epsilon_1(\omega)}{2}} \tag{30}$$

$$R(\omega) = \frac{(n-1)^2 + k^2}{(n+1)^2 + k^2} \tag{31}$$

$$\alpha(\omega) = \frac{2k\omega}{c} \tag{32}$$

$$L(\omega) = Im\left(\frac{-1}{\varepsilon(\omega)}\right) = \frac{\epsilon_2(\omega)}{\epsilon_1^2(\omega) + \epsilon_2^2(\omega)} \tag{33}$$

$$\sigma(\omega) = \sigma_1(\omega) + i\sigma_2(\omega) \tag{34}$$

In this section, we specifically examine how $LaMg_2H_7$ compounds respond to incident electric fields applied along the [100], and [001] directions.

From this, it is clear how a material reacts to incoming electromagnetic waves by looking at its dielectric function, ε(ω), which is actually quite tied to its electronic band structure. For $LaMg_2H_7$ compound studied here, both the real and imaginary parts of this function, are shown in **Figure 6(a)** in both crystallographic directions. The real part primarily describes phenomena such as polarization and dispersion effects, while the imaginary part indicates how the wave loses energy as it passes through the material. Interestingly, both compounds appear to be optically isotropic, since there's hardly any noticeable change in the imaginary part. It is observed from the real part that it starts to drop off after approximately 3.5 eV, and the imaginary part shows a similar decline, starting a bit later, around 5.5 eV.

The absorption coefficient α(ω) provides information about the optimal solar energy conversion efficiency and indicates how much light of a specific energy (wavelength) may flow through a material before being absorbed. **Figure 6(b)** illustrates that absorption begins at approximately 3.5 eV, which is consistent with the band structure and suggests that the compound exhibits nonmetallic behavior. After receiving only that amount (3.5 eV) of energy, electrons can travel to the conduction band. This compound absorbs the majority of UV radiation between 10 and 12 eV.

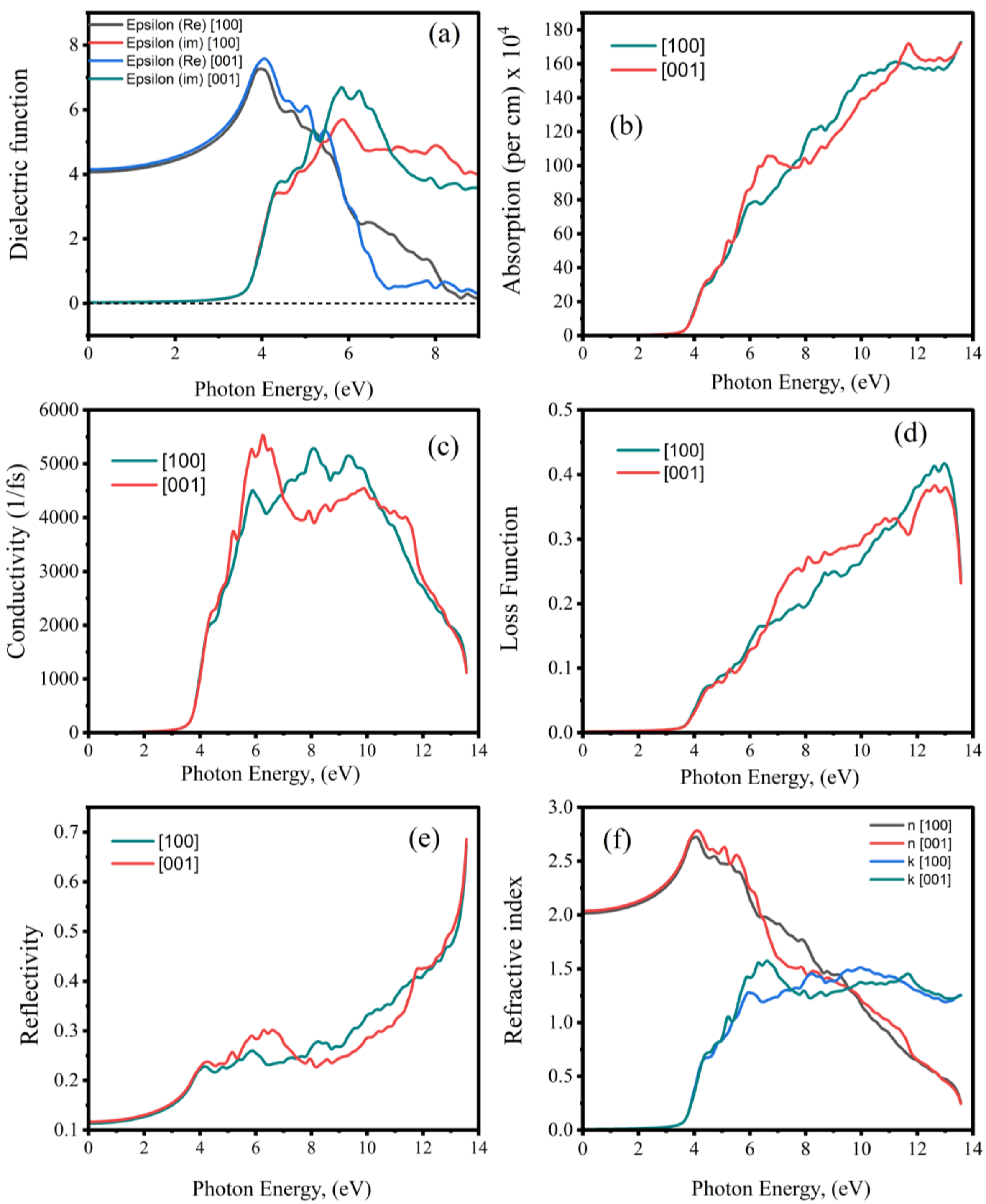

**Figure 6.** The energy-dependent (a) dielectric function, (b) absorption coefficient, (c) optical conductivity, , (d) loss function, (e) reflectivity, and (f) refractive index of $LaMg_2H_7$.

Optical conductivity, σ(ω), helps describe how free charge carriers move when exposed to specific ranges of photon energy. Since the material has a wide band gap, as shown in the band structure, photoconductivity doesn't really kick in until the photon energy reaches around 3.5 eV. This is evident in **Figure 6(c)**, which aligns well with our observations in the band structure and DOS results. Once the incoming photon energy exceeds 3.5 eV, the materials begin to absorb photons. This absorption leads to more photoconductivity, which in turn boosts their overall electrical conductivity.

**Figure 6(d)** presents the loss function L(ω) for $LaMg_2H_7$ compound. When a fast-moving electron passes through a material, it can excite collective charge oscillations, which causes the electron to lose energy. This energy loss shows up as peaks in the loss function L(ω) [62,63]. In the material, it is clearly seen that strong loss peaks are around 13 eV. These peaks mark the plasmon energy, basically, the energy at which collective electron oscillations (or plasma oscillations) occur, as described by the jellium model. Interestingly, these plasmon resonances also coincide with sudden drops in reflectivity and absorption, underscoring their physical significance. The ratio of the wave's energy reflected off the surface to the wave's energy striking the surface is known as the reflectivity, R(ω). The reflectivity for $LaMg_2H_7$ compound is displayed in **Figure 6(e)**. This compound has a reflectance of about 12% in the infrared range and more than 60% at 13 to 14 eV. So, UV radiation can be moderately reflected by these materials. Regarding the energy loss function L(ω), which describes the energy lost by fast electrons traversing the material , the main peak appears at approximately 7.5 eV in the [100] direction and 7.2 eV in the [001] direction . Based on the integration of the L(ω) curve, it is estimated that about 28–32% of the total energy loss in the 0–14 eV range is concentrated in the plasmon resonance region (approximately 6.5–8.5 eV), indicating pronounced plasmonic behavior in $LaMg_2H_7$[64].

The imaginary component of the refractive index, denoted ask, represents the attenuation experienced by an electromagnetic wave as it propagates through a medium. In contrast, the real part of the refractive index reflects the phase velocity of the wave at various energy levels. As illustrated in **Figure 6(f),** both compounds demonstrate distinct behaviors in their real and imaginary parts. Notably, the extinction coefficient shows a gradual increase from 1 eV to 4.2 eV, after which it begins to decline. Furthermore, the real part of the refractive index for both compounds exhibits an isotropic nature, with a notable decrease commencing at 3.9 eV. While the real part of the refractive index η indicates how quickly the electromagnetic wave travels through the medium at different energy levels, the imaginary part *k* provides insight into the wave's attenuation during its passage. **Figure 6(f)** presents a comprehensive view of both the imaginary and real components for the compounds across different crystallographic directions. An interesting observation is that from 3.5 eV to 6.5 eV, the extinction coefficient steadily increases, suggesting enhanced absorption characteristics in that energy range. Additionally, the real part of the refractive index exhibits slight anisotropic behavior before declining at 3.5 eV, indicating varying phase velocities in different crystallographic orientations.

## 2.5 Phonon dispersion

Calculating the phonon dispersion and phonon density of states for crystalline materials is a crucial part of modern research. These analyses provide valuable insights into the material's behavior, including its dynamic stability or instability, potential phase changes, and the impact of vibrations on properties such as heat conduction, thermal expansion, superconducting transition temperature ($T_c$), Helmholtz free energy, and heat capacity [65–67]. Here GGA(PBE-sol) functional was applied to compute the phonon dispersion along the high-symmetry directions of the Brillouin zone (Z-A-M-G-Z-R-X-G) for $LaMg_2H_7$ compound. Density functional perturbation theory was combined with a finite displacement approach to perform these calculations [68]. Phonon behavior and electron-phonon interactions play a key role in the superconducting properties of a material at temperatures below $T_c$. Assessing dynamic stability is also vital for understanding how a material will respond to mechanical stresses that change over time. **Figure 7(a)** shows the phonon dispersion curves and density of states for $LaMg_2H_7$ along the chosen high-symmetry paths in **Figure 7(b)**. Importantly, none of the phonon branches have negative frequencies, which confirms $LaMg_2H_7$ is dynamically stable in its ground-state structures.

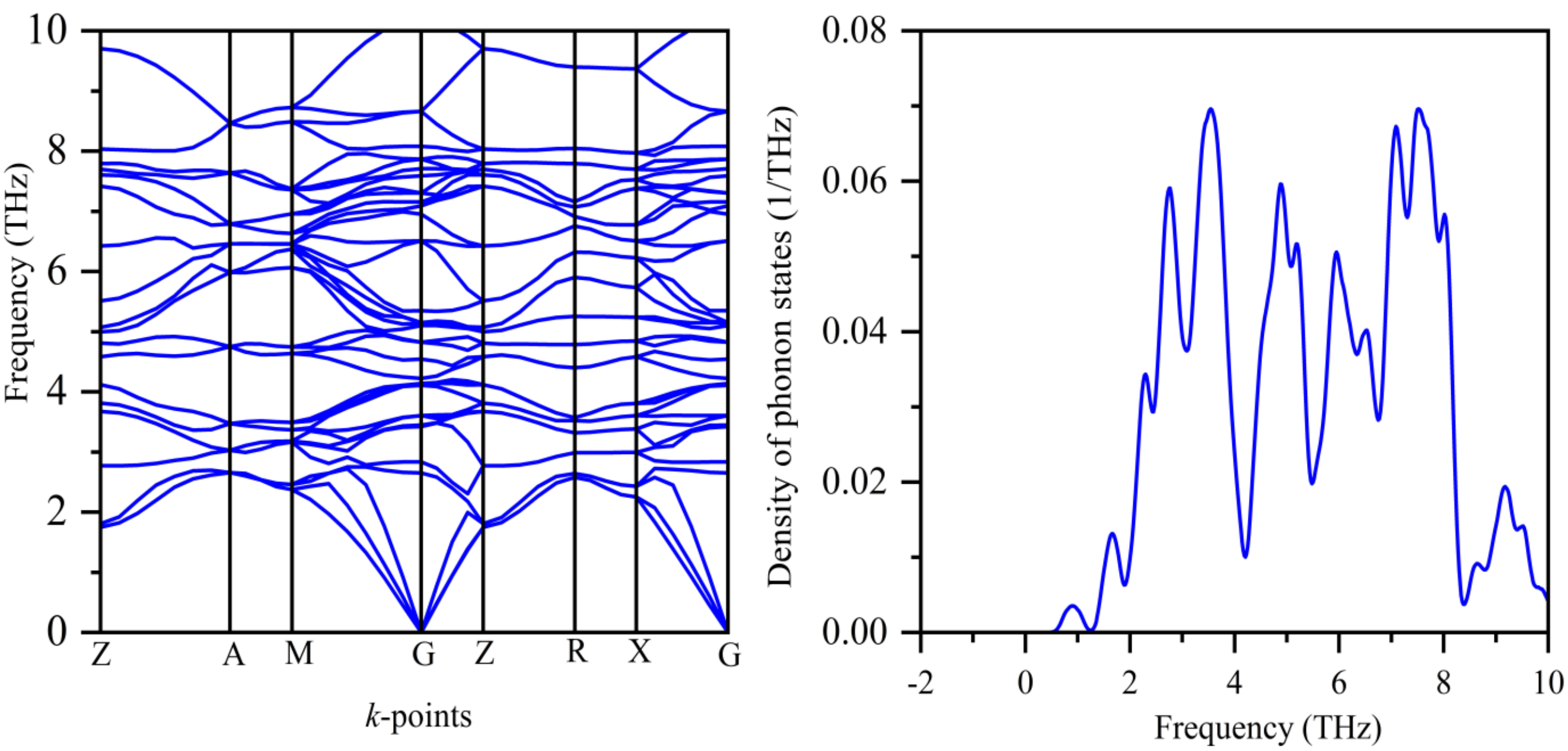

**Figure 7.** (a) Phonon dispersion and (b) phonon density of states

$LaMg_2H_7$ compound has a Tetragonal unit cell containing 40 atoms, which leads to a total of 120 vibrational modes, 3 acoustic, and 117 optical. The acoustic modes consist of one longitudinal and two transverse branches. These acoustic phonons come from atoms moving together in a regular pattern around their equilibrium positions and are closely related to sound propagation and the stiffness of the crystal. In contrast, optical phonons occur when atoms vibrate against each other, with one atom moving to the left while its neighbor moves to the right, creating higher-frequency vibrations. Acoustic phonon branches appear at lower frequencies and are mostly tied to heavier atoms in the structure, whereas optical modes appear at much higher frequencies due to the vibrations of lighter atoms. The highest-energy phonon branches originate mainly from vibrations of the lighter Mg atoms. The highest vibrational frequencies at the Γ point are 38 THz for $LaMg_2H_7$. To gain deeper insight into their lattice dynamics, **Figure 7(b)** also presents the total and partial phonon density of states (PHDOS), as well as the phonon dispersion curves. Peaks in the PHDOS appear around 3 THz and 7.5 THz for $LaMg_2H_7$ where flatter bands produce taller peaks, and more dispersive bands make the PHDOS smoother. Finally, the acoustic and optical branches overlap completely in the compound; that's why no phononic band gap is observed.

## 2.6 Thermodynamic properties

Using a quasi-harmonic approximation, the thermodynamic characteristics of $LaMg_2H_7$ are assessed in the temperature range of 200–900 K and the pressure range of 0–30 GPa. A material's resistance to uniform compression is determined by its bulk modulus. Additionally, it gives information about the material's bonding strength. The general formula ( $B = v\frac{\Delta p}{\Delta v}$ ) is satisfied by **Figure 8(a)**, which shows that the bulk modulus of $LaMg_2H_7$ rises with increasing pressure. The temperature and pressure dependency of $LaMg_2H_7$'s isothermal bulk modulus is shown in **Figure 8(b)**. The obtained results indicate that the bulk modulus of $LaMg_2H_7$ gradually decreases at temperatures over 200 K**.**

The energy content of a material resulting from the activated degrees of freedom within it is known as the internal energy, or U. **Figures 8(c)** and **8(d)** show how pressure and temperature affect $LaMg_2H_7$ internal energy. As a result, $LaMg_2H_7$ internal energy increases approximately linearly with temperature above 200 K. As pressure increases, $LaMg_2H_7$ internal energy likewise increases nearly linearly, as seen in **Figure 8c**.

Understanding the specific heat capacity of a material extends far beyond its practical importance; it also reveals a great deal about how the atoms within it vibrate and interact. At low temperatures, the heat capacity at constant volume increases roughly with the cube of the temperature, a trend predicted by the Debye model. As the temperature continues to rise, $C_v$ eventually flattens out and reaches the Dulong–Petit limit, a well-known high-temperature behavior characteristic of solids [69].

In the case of $LaMg_2H_7$**Figures 8(e)** and **8(f)** show how both $C_v$ (the specific heat at constant volume) respond to changes in pressure and temperature. When no external pressure is applied, the heat capacity increases with temperature, as shown in **Figure 8(f)**. This increase is mainly due to the softening of phonons, which essentially means that the vibrations in the atomic lattice become easier as thermal energy builds up. Up to about 300 K, this rise is particularly steep, a result of anharmonic effects that aren't captured in the simplest version of the Debye theory. On the other hand, when pressure increases while the temperature is held at 300 K, a different trend appears. As shown in **Figure 8e**, $C_v$ drops gradually. This is because added pressure restricts atomic vibrations, effectively stiffening the structure.

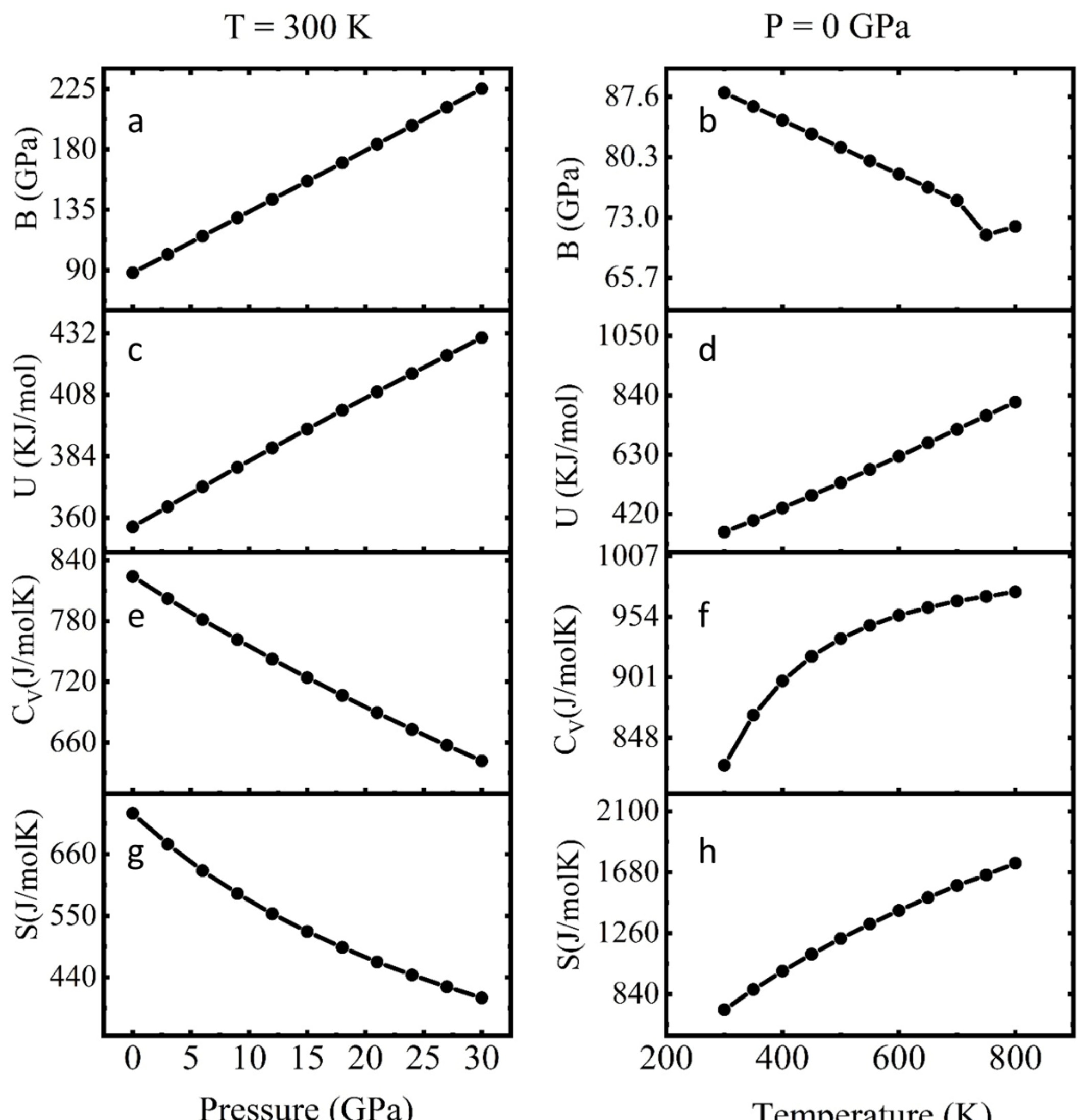

**Figure 8**. Thermodynamic properties of $LaMg_2H_7$.

An essential component of a thermodynamic system that gauges the degree of disorder present in a substance is the entropy, or S. **Figures 8(g)** and **8(h)** shows how pressure and temperature affect the change in entropy, S. Increased thermal disorder causes an increase in entropy as the temperature rises **Figure 8(h)**. Furthermore, **Figure 8(g)** shows that for T = 300 K, the entropy decreases as pressure increases.

The point at which every vibrational mode in a solid, regardless of frequency, becomes fully excited is known as the Debye temperature ($\Theta^D$). **Figures 9(a)** and **(b)** show how $\Theta^D$ changes with pressure and temperature. The value of $\Theta^D$ steadily falls with increasing temperature, exhibiting a pattern resembling that of the bulk modulus. This is because as the temperature rises, the material's stiffness decreases, resulting in a corresponding drop in both B and $\Theta^D$ [70]. However, when pressure is applied, the material becomes stiffer, which raises both B and $\Theta^D$. There is a nonlinear tendency between linear and exponential growth of $\Theta^D$ with pressure.

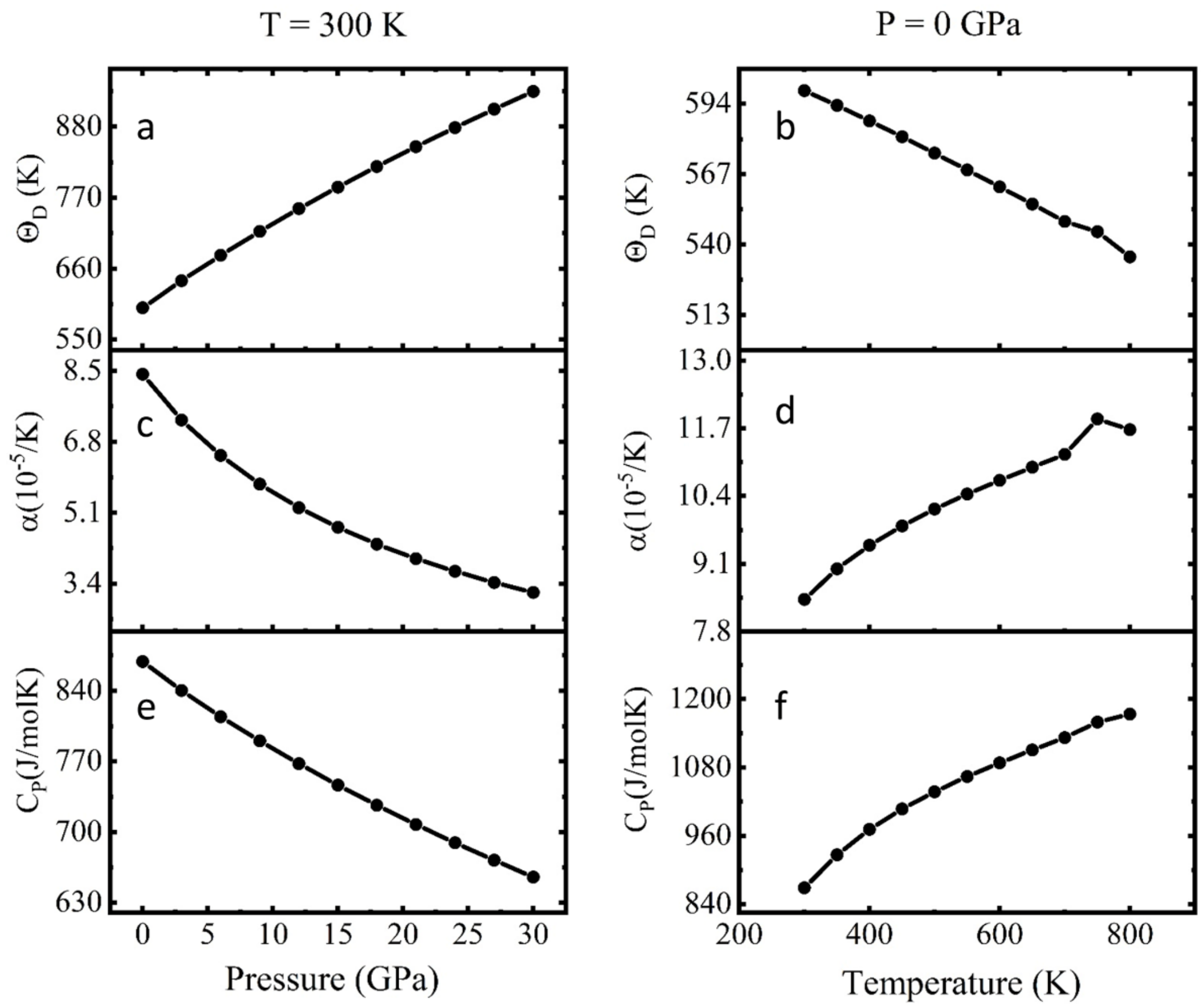

**Figure 9**. Thermodynamic properties of $LaMg_2H_7$.

The volume thermal expansion coefficient (VTEC) in relation to pressure and temperature is shown in **Figures 9(c)** and **(d)**, respectively. The coefficients of the $LaMg_2H_7$ under study decreases as pressure increases. On the other hand, at a constant pressure, the expansion coefficient increases with increasing temperature. The volume thermal expansion coefficient of a material and its bulk modulus have been shown to be inversely related. The quantity of heat needed to raise the temperature of a unit mass of a substance by 1°C (or 1 K) while keeping the pressure constant is known as the specific heat capacity at constant pressure, or $C_p$. In **Figure 9e**, $C_p$ decreases as pressure increases, while as temperature rises (**Figure 9(f)**), $C_p$ takes an uptrend.

## 2.7 Thermophysical properties

### *2.7.1 Debye temperature and sound velocity*

Among other thermophysical characteristics, a solid's Debye temperature ($\Theta_D$) dictates its bonding forces, energy of vacancy formation, melting point, thermal conductivity, phonon dynamics, specific heat, and superconductivity [71]. The Debye temperature is the temperature at which a solid's phonon wavelengths approximately correspond to the average interatomic distance. This temperature can be used to separate the lattice dynamics into high- and low-temperature zones. The Debye temperature also distinguishes between the lattice vibration's classical and quantum characteristics [72]. The energy of all vibrational modes is about $K_B T$ when T is larger than $\Theta_D$. The higher frequency modes, however, are not excited at T < $\Theta_D$ . Calculating $\Theta_D$ can be done in a number of ways. Here the Debye temperature of $LaMg_2H_7$ was determined using **Eq. 35.**

$$\Theta_D = \frac{h}{K_B}[(\frac{3n}{4\pi})\frac{N_A\rho}{M}]^{1/3}V_m \tag{35}$$

Here n is the number of atoms in the unit cell, M is the molar mass, ρ is the density, $N_A$ is Avogadro's number, $V_m$ is the mean sound velocity, h is Planck's constant, and $K_B$ is the Boltzmann constant. A high $\Theta_D$ means atoms vibrate at high characteristic frequencies, usually implying stiff interatomic bonds. Materials with high $\Theta_D$ often have high sound velocity, high hardness, and lower specific heat at low temperatures. Here high Debye temperature (≈592 K) and moderate Grüneisen (γ = 1.32) agree with stiff bonds and relatively low anharmonic scattering both favor high phonon velocities and therefore high $K_{ph}$.

One crucial factor in comprehending the thermal and acoustic characteristics of a material is the sound velocity through it. The average sound velocity in solids, $V_m$, is related to the shear modulus, bulk modulus, and crystal density [73]. The $V_m$ is given by the harmonic mean of the average longitudinal and transverse sound velocities, $V_l$ and $V_t$. The relevant relations are given in **Eqs. 36-38.**

$$V_m = [\frac{1}{3}(\frac{1}{V_l^3} + \frac{2}{V_t^3})]^{-1/3} \tag{36}$$

$$V_l = [\frac{3B+4G}{3\rho}]^{1/2} \tag{37}$$

$$V_t = [\frac{G}{\rho}]^{1/2} \tag{38}$$

**Table 7** exhibits the calculated crystal density and the acoustic velocities in $LaMg_2H_7$. Atoms move parallel to the wave propagation direction. Mainly bulk modulus (resistance to compression) and density. Higher value of $v_l$ refers material is more resistant to compression. For Transverse Sound Velocity ($v_t$), the speed of shear waves (S-waves), where atoms oscillate perpendicular to the wave direction. The shear modulus (resistance to shape change) and density. It Measures how well the material resists shearing forces. Which also Related to mechanical hardness and elastic anisotropy. For Average Sound Velocity ($v_m$) it is important for predicting heat capacity at low temperatures, phonon-mediated superconductivity, and minimum thermal conductivity. Also used in thermoelastic and ultrasonic studies.

**Table 7**. Calculated crystal density (ρ in $gm/cm^3$), longitudinal, transverse, and average sound velocities ($v_t$, $v_l$ and $v_m$ in km/s), melting temperature ($T_m$ in K), Debye temperature ($\theta_D$ in K), Grüneisen parameter ($\gamma$), thermal conductivity ($k_{ph}$ in *W/m.K*), thermal expansion coefficient (α in $K^{-1}$ ), and minimum thermal conductivity ($k_{min}$ in *W/m.K*) of $LaMg_2H_7$.

| Compound | $\rho$ | $v_t$ | $v_l$ | $v_m$ | $T_m$ | $\theta_D$ | $\gamma$ | $k_{ph}$ | $\alpha\times 10^{-5}$ | $k_{min}$ | Ref. |
|---|---|---|---|---|---|---|---|---|---|---|---|
| $LaMg_2H_7$ | 2.95 | 4.09 | 6.75 | 4.52 | 944.03 | 591.82 | 1.32 | 150.78 | 3.24 | 0.116 | This work |

### 2.7.2 *The melting temperature*

The melting temperature ($T_m$) of a solid is one of the key properties that establishes the temperature range across which it can be employed. A solid with a high melting temperature has a low coefficient of thermal expansion, high cohesive energy, and high bonding energy [74]. The melting temperature $T_m$ of solids can be found using the elastic constants and can determine by using **Eq. 39** [75].

$$T_m = \left[553K + \left(5.91\frac{K}{GPa}\right)C_{11}\right] \tag{39}$$

### 2.7.3 *Lattice thermal conductivity*

Both electrons and phonons in materials are capable of carrying thermal energy. In metals, electrons are the primary heat carriers at low temperatures. The lattice contribution increases in importance at high temperatures. Determining a material's lattice thermal conductivity is essential for high-temperature applications. When there is a temperature differential, a material's lattice thermal conductivity ($K_{ph}$) controls how much heat energy is transported via lattice vibration [76]. The $K_{ph}$ as a function of temperature can be estimated using **Eq. 40**, which was developed by Slack [75].

$$K_{ph}(T) = A(\gamma)\frac{M_{av}{\Theta_D}^3\delta}{\gamma^2 n^{\frac{2}{3}}T} \tag{40}$$

In this equation, $M_{av}$ is the average atomic mass in kg/mol, $\Theta_D$ is the Debye temperature in K, δ is the cubic root of average atomic volume in meters, n refers to the number of atoms in the conventional unit cell, T is the absolute temperature in K, and $\gamma$ is the acoustic Grüneisen parameter which determines the degree of anharmonicity of phonons. **Eqs. 41** & **42** can be used to obtain the dimensionless Grüneisen parameter from the Poisson's ratio [77].

$$\gamma = \frac{3(1+\nu)}{2(2-3\nu)} \tag{41}$$

$$A(\gamma) = \frac{5.720 \times 10^5 \times 0.849}{2 \times (1 - \frac{0.514}{\gamma} + \frac{0.228}{\gamma^2})} \tag{42}$$

The Grüneisen parameter is a dimensionless quantity that reflects the degree of lattice anharmonicity in a material. A higher value of γ indicates stronger anharmonic effects. It provides insight into how phonon frequencies and damping vary with temperature, as well as how these frequencies respond to changes in volume caused by lattice potential anharmonicity. This parameter is also important for understanding thermal expansion in crystals and is widely used to study phase transitions involving volume changes. This is a moderate value, meaning the lattice as low-to-moderate anharmonic effects. Which is  expected to  good thermal stability and moderate thermal conductivity.

### 2.7.4 Minimum thermal conductivity

The limit of a basic thermal property is the lowest thermal conductivity. High temperatures above the Debye point are required to obtain the minimum thermal conductivity ($K_{min}$), a minimal value for a compound's thermal conductivity. Importantly, the minimum thermal conductivity is independent of crystal defects like dislocations, individual vacancies, and long-range strain fields related to impurity inclusions and dislocations. To determine the lowest thermal conductivity ($K_{min}$), Clarke employed the Debye model of compounds at high temperatures to derive **Eq. 43** [78]**,** $K_B$ is the Boltzmann constant, $v_m$ is the average sound velocity and $V_{atomic}$ is the cell volume per atom.

$$K_{min} = K_B v_m \, (V_{atomic})^{-2/3} \tag{43}$$

### 2.7.5 Thermal expansion coefficient

Another noteworthy characteristic of materials is their thermal expansion coefficient (TEC). The ceramics industry uses materials with low thermal expansion [79–81]. **Eq. 44** can be used to determine the TEC from a material's shear modulus, G (in GPa) [82].

$$\alpha = \frac{1.6 \times 10^{-3}}{G} \tag{44}$$

The value of the thermal expansion coefficient is given in **Table 7**, from which it is clear that the melting temperature is moderate. For a TBC material, the extremely low minimum thermal conductivity is appropriate [83–85], and the value of the low thermal expansion coefficient is also responsible for TBC application[86]. Therefore, below about 900 K, $LaMg_2H_7$ has the potential to be utilized as TBC.

# 3 Conclusion

This study uses the density functional theory to offer a thorough first-principles analysis of the $LaMg_2H_7$ molecule. Most of the results presented here are new. Fair agreements were identified when we compared this study's findings with earlier findings in cases where they were available. The $LaMg_2H_7$ compound exhibits anisotropic character and is elastically stable. The $LaMg_2H_7$ compound shows brittleness. The hardness of this compound falls within the moderate category. The electronic band structure and the electronic energy density of states indicate it as a wide bandgap semiconductor. The effective masses and carrier concentrations have also been explored. The thermodynamic properties have also been discussed at varying pressures and temperatures. $LaMg_2H_7$ exhibits a moderately high melting point along with a low minimum thermal conductivity. This material is a promising candidate for use in hydrogen fuel cell applications. $LaMg_2H_7$ It can be used as a moderate-level reflector for UV rays and is also a perfect material for TBC applications. Finally, we have presented several new results on a variety of properties of $LaMg_2H_7$. The results of this study should inspire materials scientists to investigate this intriguing system further using both theoretical and experimental methods.

**Author contribution**

**Tanvir Khan & Md H. S. Rifat:** Investigation, Conceptualization, Data curation, Formal analysis, Visualization, Writing - original draft. **M. Ibrahim:** methodology and software. **J.H. Abir:** methodology, software and Writing. methodology, software, **S. Roy:** methodology, and software, **S.H. Naqib:** Supervision, Formal Analysis, Writing-Reviewing and Editing.

# References

1. Abe, J. O., Popoola, A. P. I., Ajenifuja, E. & Popoola, O. M. Hydrogen energy, economy and storage: Review and recommendation. *Int. J. Hydrogen Energy* **44**, 15072–15086 (2019).
2. Sadek, O. *et al.* Synthesis by sol-gel method and characterization of nano-$TiO_2$ powders. *Mater. Today Proc.* **66**, 456–458 (2022).
3. Anoua, R. *et al.* Optical and electronic properties of the natural Alizarin dye: Theoretical and experimental investigations for DSSCs application. *Opt. Mater. (Amst).* **127**, 112113 (2022).
4. Hanley, E. S., Deane, J. & Gallachóir, B. Ó. The role of hydrogen in low carbon energy futures–A review of existing perspectives. *Renew. Sustain. Energy Rev.* **82**, 3027–3045 (2018).
5. Ley, M. B. *et al.* Complex hydrides for hydrogen storage – new perspectives. *Mater. Today* **17**, 122–128 (2014).
6. Lehrbuch der Kristallphysik.
7. Jain, I. P., Lal, C. & Jain, A. Hydrogen storage in Mg: A most promising material. *Int. J. Hydrogen Energy* **35**, 5133–5144 (2010).
8. Schlapbach, L. & Züttel, A. Hydrogen-storage materials for mobile applications. *Nature* **414**, 353–358 (2001).
9. Barkhordarian, G., Klassen, T. & Bormann, R. Fast hydrogen sorption kinetics of nanocrystalline Mg using $Nb_2O_5$ as catalyst. *Scr. Mater.* **49**, 213–217 (2003).
10. Zaluska, A., Zaluski, L. & Ström–Olsen, J. . Nanocrystalline magnesium for hydrogen storage. *J. Alloys Compd.* **288**, 217–225 (1999).
11. Hirscher, M. *et al.* Materials for hydrogen-based energy storage - past, recent progress and future outlook. *J. Alloys Compd.* **827**, 153548 (2020).
12. Dai, J. H., Song, Y. & Yang, R. First Principles Study on Hydrogen Desorption from a Metal (=Al, Ti, Mn, Ni) Doped $MgH_2$ (110) Surface. *J. Phys. Chem. C* **114**, 11328–11334 (2010).
13. Yildirim, T. & Ciraci, S. Titanium-Decorated Carbon Nanotubes as a Potential High-Capacity Hydrogen Storage Medium. *Phys. Rev. Lett.* **94**, 175501 (2005).
14. Ströbel, R., Garche, J., Moseley, P. T., Jörissen, L. & Wolf, G. Hydrogen storage by carbon materials. *J. Power Sources* **159**, 781–801 (2006).
15. Dillon, A. C. *et al.* Storage of hydrogen in single-walled carbon nanotubes. *Nature* **386**, 377–379 (1997).
16. Chambers, A., Park, C., Baker, R. T. K. & Rodriguez, N. M. Hydrogen Storage in Graphite Nanofibers. *J. Phys. Chem. B* **102**, 4253–4256 (1998).
17. Chen, C.-H. & Huang, C.-C. Enhancement of hydrogen spillover onto carbon nanotubes with defect feature. *Microporous Mesoporous Mater.* **109**, 549–559 (2008).
18. Bertheville, B., Fischer, P. & Yvon, K. High-pressure synthesis and crystal structures of new ternary caesium magnesium hydrides, CsMgH3, Cs4Mg3H10 and $Cs_2MgH_4$. *J. Alloys Compd.* **330–332**, 152–156 (2002).
19. Gingla, F., Vogt, T., Akiba, E. & Yvon, K. Cubic CsCaH3 and hexagonal RbMgH3: new examples of

fluoride-related perovskite-type hydrides. *J. Alloys Compd.* **282**, 125–129 (1999).

20. Blaha, P. *et al.* WIEN2k: An APW+lo program for calculating the properties of solids. *J. Chem. Phys.* **152**, 74101 (2020).
21. McCabe, C. J., Halvorson, M. A., King, K. M., Cao, X. & Kim, D. S. Interpreting Interaction Effects in Generalized Linear Models of Nonlinear Probabilities and Counts. *Multivariate Behav. Res.* **57**, 243–263 (2022).
22. Perdew, J. P. *et al.* Restoring the Density-Gradient Expansion for Exchange in Solids and Surfaces. *Phys. Rev. Lett.* **100**, 136406 (2008).
23. Jiang, H. Band gaps from the Tran-Blaha modified Becke-Johnson approach: A systematic investigation. *J. Chem. Phys.* **138**, (2013).
24. Monkhorst, H. J. & Pack, J. D. Special points for Brillouin-zone integrations. *Phys. Rev. B* **13**, 5188–5192 (1976).
25. Otero-de-la-Roza, A., Abbasi-Pérez, D. & Luaña, V. Gibbs2: A new version of the quasiharmonic model code. II. Models for solid-state thermodynamics, features and implementation. *Comput. Phys. Commun.* **182**, 2232–2248 (2011).
26. Gingl, F., Yvon, K., Vogt, T. & Hewat, A. Synthesis and crystal structure of tetragonal LnMg2H7 (Ln=La, Ce), two Laves phase hydride derivatives having ordered hydrogen distribution. *J. Alloys Compd.* **253–254**, 313–317 (1997).
27. Surucu, G., Candan, A., Gencer, A. & Isik, M. First-principle investigation for the hydrogen storage properties of NaXH3 (X= Mn, Fe, Co) perovskite type hydrides. *Int. J. Hydrogen Energy* **44**, 30218–30225 (2019).
28. Wang, M. *et al.* Reversible calcium alloying enables a practical room-temperature rechargeable calcium-ion battery with a high discharge voltage. *Nat. Chem.* **10**, 667–672 (2018).
29. Chaib, H. *et al.* Effect of metal atom substitutions in Li based hydrides for hydrogen storage. *Int. J. Hydrogen Energy* **45**, 28920–28929 (2020).
30. Rizwan, M. *et al.* Effect of electronic alteration on hydrogen storage and optical response in NaMgF3 using DFT approach. *Int. J. Hydrogen Energy* **48**, 33599–33609 (2023).
31. Pan, Y. & Guan, W. M. The hydrogenation mechanism of PtAl and IrAl thermal barrier coatings from first-principles investigations. *Int. J. Hydrogen Energy* **45**, 20032–20041 (2020).
32. Pan, Y., Lin, Y., Liu, G. & Zhang, J. Influence of transition metal on the mechanical and thermodynamic properties of IrAl thermal barrier coating. *Vacuum* **174**, 109203 (2020).
33. Pan, Y., Pu, D., Liu, G. & Wang, P. Influence of alloying elements on the structural stability, elastic, hardness and thermodynamic properties of Mo5SiB2 from first-principles calculations. *Ceram. Int.* **46**, 16605–16611 (2020).
34. Pu, D. L. & Pan, Y. Influence of high pressure on the structure, hardness and brittle-to-ductile transition of NbSi2 ceramics. *Ceram. Int.* **47**, 2311–2318 (2021).
35. Lord, E. A. The Dirac spinor in six dimensions. *Math. Proc. Cambridge Philos. Soc.* **64**, 765–778 (1968).
36. Schöllhammer, G. & Herzig, P. Elastic constants of La, LaH2, and LaH3. *Monatshefte für Chemie - Chem. Mon.* **143**, 1325–1328 (2012).
37. Yamçıçıer, Ç. Exploring the structural, elastic, phonon, optoelectronics, and thermoelectric properties of tetragonal complex metal hydride X2MgH4 (X=K, Rb, and Cs) compounds for hydrogen storage applications. *Int. J. Hydrogen Energy* **48**, 39930–39943 (2023).
38. Voigt, W. Ueber die Beziehung zwischen den beiden Elasticitätsconstanten isotroper Körper. *Ann. Phys.* **274**, 573–587 (1889).
39. Reuss, A. Berechnung der Fließgrenze von Mischkristallen auf Grund der Plastizitätsbedingung für Einkristalle . *ZAMM - J. Appl. Math. Mech. / Zeitschrift für Angew. Math. und Mech.* **9**, 49–58 (1929).
40. Pugh, S. F. XCII. Relations between the elastic moduli and the plastic properties of polycrystalline pure metals. *London, Edinburgh, Dublin Philos. Mag. J. Sci.* **45**, 823–843 (1954).

41. Ali, M. A. *et al.* Recently synthesized (Zr1-xTix)2AlC ($0 \leq x \leq 1$) solid solutions: Theoretical study of the effects of M mixing on physical properties. *J. Alloys Compd.* **743**, 146–154 (2018).

42. Ranganathan, S. I. & Ostoja-Starzewski, M. Universal Elastic Anisotropy Index. *Phys. Rev. Lett.* **101**, 55504 (2008).

43. Meneve, J., Vercammen, K., Dekempeneer, E. & Smeets, J. Thin tribological coatings: magic or design? *Surf. Coatings Technol.* **94–95**, 476–482 (1997).

44. Liu, Z. & Scanlon, M. G. Modelling Indentation of Bread Crumb by Finite Element Analysis. *Biosyst. Eng.* **85**, 477–484 (2003).

45. Cheng, Y.-T. & Cheng, C.-M. Scaling, dimensional analysis, and indentation measurements. *Mater. Sci. Eng. R Reports* **44**, 91–149 (2004).

46. Gaillac, R., Pullumbi, P. & Coudert, F.-X. ELATE: an open-source online application for analysis and visualization of elastic tensors. *J. Phys. Condens. Matter* **28**, 275201 (2016).

47. Nørskov, J. K., Bligaard, T., Rossmeisl, J. & Christensen, C. H. Towards the computational design of solid catalysts. *Nat. Chem.* **1**, 37–46 (2009).

48. Hautier, G., Fischer, C. C., Jain, A., Mueller, T. & Ceder, G. Finding Nature's Missing Ternary Oxide Compounds Using Machine Learning and Density Functional Theory. *Chem. Mater.* **22**, 3762–3767 (2010).

49. Calandra, M., Kolmogorov, A. N. & Curtarolo, S. Search for high <math display="inline"> <msub> <mi>T</mi> <mi>c</mi> </msub> </math> in layered structures: The case of LiB. *Phys. Rev. B* **75**, 144506 (2007).

50. Neamen, D. A. & Biswas, D. *Semiconductor Physics and Devices*. (McGraw-Hill higher education New York, 2011).

51. Islam, M. K., Sarker, M. A. R., Inagaki, Y. & Islam, M. S. Study of a new layered ternary chalcogenide CuZnTe 2 and its potassium intercalation effect. *Mater. Res. Express* **7**, 105904 (2020).

52. Pazos-Outón, L. M., Xiao, T. P. & Yablonovitch, E. Fundamental Efficiency Limit of Lead Iodide Perovskite Solar Cells. *J. Phys. Chem. Lett.* **9**, 1703–1711 (2018).

53. Liu, X. *et al.* A high dielectric constant non-fullerene acceptor for efficient bulk-heterojunction organic solar cells. *J. Mater. Chem. A* **6**, 395–403 (2018).

54. Pankove, J. I. *Optical Processes in Semiconductors*. (Courier Corporation, 1975).

55. Kim, K. S. *et al.* The interface formation and adhesion of metals (Cu, Ta, and Ti) and low dielectric constant polymer-like organic thin films deposited by plasma-enhanced chemical vapor deposition using para-xylene precursor. *Thin Solid Films* **377–378**, 122–128 (2000).

56. Mott, N. F. & Davis, E. A. *Electronic Processes in Non-Crystalline Materials*. (OUP Oxford, 2012).

57. Boubaker, K. A physical explanation to the controversial Urbach tailing universality. *Eur. Phys. J. Plus* **126**, 10 (2011).

58. Choudhury, B., Borah, B. & Choudhury, A. Extending Photocatalytic Activity of TiO 2 Nanoparticles to Visible Region of Illumination by Doping of Cerium. *Photochem. Photobiol.* **88**, 257–264 (2012).

59. Mulliken, R. S. Electronic Population Analysis on LCAO–MO Molecular Wave Functions. I. *J. Chem. Phys.* **23**, 1833–1840 (1955).

60. Kohn, W. An essay on condensed matter physics in the twentieth century. *Rev. Mod. Phys.* **71**, S59–S77 (1999).

61. Liu, Q.-J., Liu, Z.-T., Feng, L.-P. & Tian, H. First-principles study of structural, elastic, electronic and optical properties of rutile GeO2 and α-quartz GeO2. *Solid State Sci.* **12**, 1748–1755 (2010).

62. Naher, M. I. & Naqib, S. H. First-principles insights into the mechanical, optoelectronic, thermophysical, and lattice dynamical properties of binary topological semimetal BaGa2. *Results Phys.* **37**, 105507 (2022).

63. Segall, M. D. *et al.* First-principles simulation: ideas, illustrations and the CASTEP code. *J. Phys. Condens. Matter* **14**, 2717–2744 (2002).

64. Ambrosch-Draxl, C. & Sofo, J. O. Linear optical properties of solids within the full-potential linearized augmented planewave method. *Comput. Phys. Commun.* **175**, 1–14 (2006).

65. Disa, A. S., Nova, T. F. & Cavalleri, A. Engineering crystal structures with light. *Nat. Phys.* **17**, 1087–1092 (2021).

66. Kolmogorov, A. N., Calandra, M. & Curtarolo, S. Thermodynamic stabilities of ternary metal borides: An ab initio guide for synthesizing layered superconductors. *Phys. Rev. B* **78**, 094520 (2008).

67. Yun, Y., Legut, D. & Oppeneer, P. M. Phonon spectrum, thermal expansion and heat capacity of UO2 from first-principles. *J. Nucl. Mater.* **426**, 109–114 (2012).

68. Kresse, G., Furthmüller, J. & Hafner, J. Ab initio Force Constant Approach to Phonon Dispersion Relations of Diamond and Graphite. *Europhys. Lett.* **32**, 729–734 (1995).

69. Burton, A. C. The Application of the Theory of Heat Flow to the Study of Energy Metabolism. *J. Nutr.* **7**, 497–533 (1934).

70. Parvin, F. & Naqib, S. H. Pressure dependence of structural, elastic, electronic, thermodynamic, and optical properties of van der Waals-type NaSn2P2 pnictide superconductor: Insights from DFT study. *Results Phys.* **21**, 103848 (2021).

71. Tang, K., Wang, T., Qi, W. & Li, Y. Debye temperature for binary alloys and its relationship with cohesive energy. *Phys. B Condens. Matter* **531**, 95–101 (2018).

72. Ravindran, P. *et al.* Density functional theory for calculation of elastic properties of orthorhombic crystals: Application to TiSi2. *J. Appl. Phys.* **84**, 4891–4904 (1998).

73. Trachenko, K., Monserrat, B., Pickard, C. J. & Brazhkin, V. V. Speed of sound from fundamental physical constants. *Sci. Adv.* **6**, (2020).

74. Fine, M. E., Brown, L. D. & Marcus, H. L. Elastic constants versus melting temperature in metals. *Scr. Metall.* **18**, 951–956 (1984).

75. Slack, G. A. The Thermal Conductivity of Nonmetallic Crystals. in 1–71 (1979). doi:10.1016/S0081-1947(08)60359-8.

76. Jaafreh, R., Kang, Y. S. & Hamad, K. Lattice Thermal Conductivity: An Accelerated Discovery Guided by Machine Learning. *ACS Appl. Mater. Interfaces* **13**, 57204–57213 (2021).

77. Julian, C. L. Theory of Heat Conduction in Rare-Gas Crystals. *Phys. Rev.* **137**, A128–A137 (1965).

78. Clarke, D. R. Materials selection guidelines for low thermal conductivity thermal barrier coatings. *Surf. Coatings Technol.* **163–164**, 67–74 (2003).

79. Sowa, H., Macavei, J. & Schultz, H. The crystal structure of berlinite AlPO 4 at high pressure. *Zeitschrift für Krist. - Cryst. Mater.* **192**, 119–136 (1990).

80. Nørskov, J. K., Bligaard, T., Rossmeisl, J. & Christensen, C. H. Towards the computational design of solid catalysts. *Nat. Chem.* **1**, 37–46 (2009).

81. Kerdsongpanya, S., Alling, B. & Eklund, P. Effect of point defects on the electronic density of states of ScN studied by first-principles calculations and implications for thermoelectric properties. *Phys. Rev. B* **86**, 195140 (2012).

82. Naher, M. I. & Naqib, S. H. An ab-initio study on structural, elastic, electronic, bonding, thermal, and optical properties of topological Weyl semimetal TaX (X = P, As). *Sci. Rep.* **11**, 5592 (2021).

83. Ali, M. A. *et al.* Recently synthesized (Zr1-xTix)2AlC ($0 \leq x \leq 1$) solid solutions: Theoretical study of the effects of M mixing on physical properties. *J. Alloys Compd.* **743**, 146–154 (2018).

84. Mridha, M. M. & Naqib, S. H. Pressure dependent elastic, electronic, superconducting, and optical properties of ternary barium phosphides (Ba M 2 P 2 ; M = Ni, Rh): DFT based insights. *Phys. Scr.* **95**, 105809 (2020).

85. Naher, M. I., Parvin, F., Islam, A. K. M. A. & Naqib, S. H. Physical properties of niobium-based intermetallics (Nb3B; B = Os, Pt, Au): a DFT-based ab-initio study. *Eur. Phys. J. B* **91**, 289 (2018).

86. Reza, A. S. M. M., Afzal, M. A. & Naqib, S. H. First-principles investigation of the physical properties of wide band gap hexagonal AlPO4 compound for possible applications. *Mater. Sci. Semicond. Process.* **190**,

109322 (2025).